\documentclass[reprint,superscriptaddress,showpacs,amsmath,amssymb,aps,pra]{revtex4-1}

\usepackage{graphicx}
\usepackage{dcolumn}
\usepackage{bm}
\usepackage{epsfig}
\usepackage{subfigure}

\begin{document}

\title{Frictionless atom cooling in harmonic traps:\\a time-optimal approach}

\author{Dionisis Stefanatos}
\email{dionisis@seas.wustl.edu}
\author{Justin Ruths}
\author{Jr-Shin Li}
\affiliation{Department of Electrical and Systems Engineering, Washington University, St. Louis, MO 63130, USA}

\date{\today}

\begin{abstract}

In this article we formulate frictionless atom cooling in harmonic traps as a time-optimal control problem, permitting imaginary values of the trap frequency for trasient time intervals during which the trap becomes an expulsive parabolic potential. We show that the minimum time solution has ``bang-bang" form, where the frequency jumps suddenly at certain instants and then remains constant, and calculate estimates of the minimum cooling time for various numbers of such jumps. A numerical optimization method based on pseudospectral approximations is used to obtain suboptimal realistic solutions without discontinuities, which may be implemented experimentally.

\end{abstract}

\maketitle

\section{Introduction}

Frictionless atom cooling in a harmonic trap is defined as the problem of changing the harmonic frequency of the trap to some lower final value, while keeping the populations of the initial and final levels invariant, thus without generating friction and heating. Achieving this goal in minimum time has many important potential applications. For example, it can be used to reach extremely low temperatures inaccessible by standard cooling techniques \cite{Leanhardt03}, to reduce the velocity dispersion  and collisional shifts for spectroscopy and atomic clocks \cite{Bize05}, and in adiabatic quantum computation \cite{Aharonov07}. It is also closely related to the problem of moving in minimum time a system between two thermal states, as for example in the transition from graphite to diamond \cite{Salamon09}. By using optimal control theory, it was proved that minimum transfer time can be achieved with ``bang-bang" real frequency processes, where the frequencies change suddenly at certain instants and then stay constant \cite{Salamon09}. In another recent paper \cite{Chen10}, it was shown that when the restriction for real frequencies is relaxed, allowing the trap to become an expulsive parabolic potential at some time intervals, shorter transfer times can be obtained. Based on the theory presented in \cite{Chen10}, in this article we reformulate the frictionless cooling problem as a minimum-time optimal control problem, permitting the frequency to take real and imaginary values in specified ranges. We then show that the optimal solution has ``bang-bang" form and use this fact to calculate estimates of the minimum transfer time for various numbers of frequency jumps. We finally use a numerical optimization method based on pseudospectral approximations to find suboptimal realistic solutions which do not suffer from discontinuities and are thus appropriate for experimental implementation. The efficiency of the method is demonstrated by several numerical examples.

\section{Formulation of the cooling problem in terms of optimal control}

Consider the one-dimensional time-dependent harmonic oscillator with Hamiltonian
\begin{equation}
\label{hoscillator}
H(t)=\frac{1}{2m}\hat{p}^2+\frac{m\omega^2(t)}{2}\hat{q}^2,
\end{equation}
with initial frequency $\omega(0)=\omega_0$ at $t=0$ and final frequency $\omega(t_f)=\omega_f<\omega_0$ at the final time $t_f$. This corresponds to a temperature reduction by a factor $\omega_f/\omega_0$. The goal is to find a path $\omega(t)$ between these two values so that the populations of all the oscillator levels $n=0,1,2,\ldots$ at $t=t_f$ are equal to the ones at $t=0$. We would also like to achieve this in minimum time $t_f$. It was shown in \cite{Chen10} that appropriate $\omega(t)$ can be efficiently engineered by using an invariant of the motion \eqref{hoscillator}. Additionally, by relaxing the restriction $\omega^2(t)\geq 0$, allowing $\omega^2(t)<0$ for some time intervals where the potential becomes expulsive, shorter cooling times can be obtained. In the following we present an overview of the corresponding theory, which will lead naturally to the formulation of the problem in terms of optimal control.

The basis of the analysis is the invariant of the motion
\begin{equation}
\label{invariant}
I(t)=\frac{m\omega_0^2}{2}\left(\frac{\hat{q}}{b}\right)^2+\frac{1}{2m}\hat{\pi}^2,
\end{equation}
where $\hat{\pi}=b\hat{p}-m\dot{b}\hat{q}$ plays the role of a momentum conjugate to $\hat{q}/b$ and the dots represent derivatives with respect to time \cite{Lewis69}. The scaling dimensionless function $b=b(t)$ satisfies the subsidiary condition
\begin{equation}
\label{Ermakov}
\ddot{b}+\omega^2(t)b=\frac{\omega_0^2}{b^3},
\end{equation}
an Ermakov equation where real solutions must be chosen to make $I$ Hermitian. $I(t)$ has the structure of a harmonic oscillator Hamiltonian, with time-dependent eigenvectors $|n(t)\rangle$ and time-independent eigenvalues $(n+1/2)\hbar\omega_0$. The general solution of the Schr\"{o}dinger equation is a superposition of orthonormal ``expanding modes"
\begin{equation}
\label{superposition}
\psi(t,x)=\sum_nc_ne^{i\alpha_n(t)}\langle x|n(t)\rangle,
\end{equation}
where $\alpha_n(t)=-(n+1/2)\omega_0\int_0^tdt'/b^2$, and $c_n$ are time-independent amplitudes. For a single mode,
\begin{multline}
\label{singlemode}
\Psi_n(t,x)=\left(\frac{m\omega_0}{\pi\hbar}\right)^{1/4}\frac{e^{i\alpha_n(t)}}{(2^nn!b)^{1/2}}\times
\\\exp{\left[i\frac{m}{2\hbar}\left(\frac{\dot{b}}{b}+\frac{i\omega_0}{b^2}\right)x^2\right]}
H_n\left[\left(\frac{m\omega_0}{\hbar}\right)^{1/2}\frac{x}{b}\right],
\end{multline}
where $H_n$ is the Hermite polynomial of degree $n$. The time-dependent average energy of the mode is
\begin{equation}
\label{energy}
\langle H(t)\rangle_n=\frac{(2n+1)\hbar}{4\omega_0}\left[\dot{b}^2+\omega^2(t)b^2+\frac{\omega_0^2}{b^2}\right].
\end{equation}
The average position is zero and the standard deviation $\sigma=(\int x^2|\Psi_n|^2dx)^{1/2}$ is proportional to $b$, $\sigma=b(n+1/2)^{1/2}/(m\omega_0/\hbar)^{1/2}$, which underlines the physical meaning of the scaling factor.

The approach taken in \cite{Chen10} is to leave $\omega(t)$ undetermined at first and impose properties on $b$ and its derivatives at the boundaries $t=0$ and $t=t_f$ to assure that:
\begin{enumerate}
\item Any eigenstate of $H(0)$ evolves as a single expanding mode
\item This expanding mode becomes, up to a position-independent phase factor, equal to the corresponding eigenstate of the Hamiltonian $H(t_f)$ of the final trap.
\end{enumerate}
When the above are satisfied, the populations in the instantaneous basis are kept equal at the initial and final times. It is not hard to find the corresponding boundary conditions. By choosing $b(0)=1, \dot{b}(0)=0$ at $t=0$, $H(0)$ and $I(0)$ commute and have common eigenfunctions at that instant. Since $\omega(0)=\omega_0$, it holds that $\ddot{b}(0)=0$ from \eqref{Ermakov}. These conditions imply that any initial eigenstate of $H(0)$ will evolve according to the expanding mode \eqref{singlemode}. In general $H(t)$ and $I(t)$ will not commute for $t>0$. At $t=t_f$ it is desirable for $\Psi_n(t_f,x)$ to be proportional, up to the global phase factor $e^{i\alpha_n(t_f)}$, to the corresponding eigenstate of the final trap. If we impose $b(t_f)=\gamma=(\omega_0/\omega_f)^{1/2}, \dot{b}(t_f)=0,\ddot{b}(t_f)=0$, then from \eqref{Ermakov} we get $\omega(t_f)=\omega_f$ and from \eqref{singlemode} we see that $\Psi_n(t_f,x)$ has the desired form. After fixing $b(t)$ and its derivatives at the boundaries, $b(t)$ can be chosen as a real function satisfying these conditions. For example, substituting the simple polynomial ansatz
\begin{equation}
\label{polynomialansatz}
b(t)=\sum_{j=0}^5a_jt^j
\end{equation}
into the six boundary conditions gives six equations that can be solved to provide the coefficients,
\begin{equation}
\label{polynomial}
b(t)=6(\gamma-1)s^5-15(\gamma-1)s^4+10(\gamma-1)s^3+1,
\end{equation}
where $s=t/t_f$. Once $b(t)$ has been determined, the physical frequency $\omega(t)$ is obtained from the subsidiary condition \eqref{Ermakov}.

Note that in the above method, the duration $t_f$ is considered to be fixed and there are no bounds on the frequency $\omega(t)$. An alternative approach is to express the frictionless cooling problem as a minimum-time optimal control problem, incorporating possible restrictions on $\omega(t)$ due for example to experimental limitations. If we set
\begin{equation}
\label{definitions}
x_1=b,\,x_2=\frac{\dot{b}}{\omega_0},\,u(t)=\frac{\omega^2(t)}{\omega_0^2},
\end{equation}
and rescale time according to $t_{\mbox{new}}=\omega_0 t_{\mbox{old}}$, we obtain the following system of first order differential equations, equivalent to the Ermakov equation \eqref{Ermakov}
\begin{eqnarray}
\label{system1}
\dot{x}_1 & = & x_2,\\
\label{system2}
\dot{x}_2 & = & -ux_1+\frac{1}{x_1^3}.
\end{eqnarray}
The optimal control problem is: Find $-v_1\leq u(t)\leq v_2$ with $u(0)=1, u(t_f)=1/\gamma^4$ such that starting from $(x_1(0),x_2(0))=(1,0)$, the above system reaches the final point $(x_1(t_f),x_2(t_f))=(\gamma,0)$ in minimum time $t_f$ (note that $\gamma=(\omega_0/\omega_f)^{1/2}>1$). The boundary conditions on the state variables $(x_1,x_2)$ are equivalent to those for $b$ and $\dot{b}$, while the boundary conditions on the control variable $u$ lead to the corresponding conditions for $\ddot{b}$. Parameters $v_1,v_2>0$ define the allowable values of $u(t)$ with $v_2\geq u(0)=1$. Note that the possibility $\omega^2(t)<0$ (expulsive parabolic potential) for some time intervals is permitted. Finally observe that the above system describes the one-dimensional newtonian motion of a unit-mass particle, with position coordinate $x_1$ and velocity $x_2$. The acceleration (force) acting on the particle is $-ux_1+1/x_1^3$. This point of view can provide useful physical insight, as we will see later.

The advantage of expressing the cooling problem in terms of optimal control is that analytical and numerical tools from this area can be used to engineer $\omega(t)$, while taking into account possible limitations on the frequency. The control-theoretical framework has been successfully employed to solve various problems in quantum dynamics \cite{Tarn80, Peirce88, Khaneja01, D'Alessandro01, Lloyd01, Sklarz02, Boscain02, Skinner03, Stefanatos04, Stefanatos05, Li06, Gorshkov08, Maximov08, Li09, Liieee09, Wu09, Lapert10, Schulte10, Stefanatos10}. We show how this can be done for the problem at hand in the following sections.

\section{Theoretical optimal solution of bang-bang type}

The form of the theoretical time-optimal solution can be found using Pontryagin's maximum principle \cite{Pontryagin}, which we state here in order to keep the paper self-contained.

\textbf{Maximum Principle for Time-Optimal Problems}: Consider the autonomous dynamical system
\begin{equation}
\label{autonomous}
\dot{\mathbf{x}}=\mathbf{f}(\mathbf{x},\mathbf{u}),
\end{equation}
where $\mathbf{x}=(x_1,x_2,\ldots,x_n)\in\mathbf{X}$ (state space), $\mathbf{u}=(u_1,u_2,\ldots,u_m)\in\mathbf{U}$ (control region), and $\mathbf{f}=(f_1,f_2,\ldots,f_n)$, with functions $f_i(\mathbf{x},\mathbf{u})$ continuous in the variables $\mathbf{x},\mathbf{u}$ and continuously differentiable with respect to $\mathbf{x}$. The corresponding control Hamiltonian is defined as
\begin{equation}
H_c(\mathbf{p},\mathbf{x},\mathbf{u})=\sum_{i=1}^n p_nf_n(\mathbf{x},\mathbf{u}),
\end{equation}
where $\mathbf{p}=(p_1,p_2,\ldots,p_n)$ is the adjoint vector. Let $\mathbf{u}(t), 0\leq t \leq t_f$, be an admissible control which transfers the state vector from $\mathbf{x}_0$ to $\mathbf{x}_f$, and let $\mathbf{x}(t)$ be the corresponding trajectory, so that $\mathbf{x}(0)=\mathbf{x}_0, \mathbf{x}(t_f)=\mathbf{x}_f$. For $\mathbf{u}(t),\mathbf{x}(t)$ to be time-optimal, it is necessary that there exists a \textbf{nonzero}, continuous vector function $\mathbf{p}(t)=(p_1(t),p_2(t),\ldots,p_n(t))$ such that:
\begin{enumerate}
\item
\begin{eqnarray}
\dot{\mathbf{x}} & = & \frac{\partial H_c}{\partial\mathbf{p}}\\
\label{adjoint}
\dot{\mathbf{p}} & = & -\frac{\partial H_c}{\partial\mathbf{x}}
\end{eqnarray}
The first equation is equivalent to the system equation (\ref{autonomous}), while the other is the equation for the adjoint vector.
\item For all $0\leq t \leq t_f$ the function $H_c(\mathbf{p}(t),\mathbf{x}(t),\mathbf{u})$ of the variable $\mathbf{u}\in\mathbf{U}$ attains its maximum at the point $\mathbf{u}=\mathbf{u}(t)$.
\item $H_c(\mathbf{p}(t),\mathbf{x}(t),\mathbf{u}(t))=c\geq 0$, $c$ constant.
\end{enumerate}

For the system described by \eqref{system1} and \eqref{system2} the states $(x_1,x_2)\in\mathbf{X}=(0,+\infty)\times R$ and the control $u\in\mathbf{U}=[-v_1,v_2]$. Note that $x_1(t)>0$ because the starting point is $x_1(0)=1>0$, the evolution is continuous for $x_1\neq 0$ and when $x_1\rightarrow 0^+$ there is a ``repulsive force" $\thicksim 1/x_1^3$ that forces $x_1$ to increase. The system equations satisfy the necessary smoothness conditions in spaces $\mathbf{X},\mathbf{U}$. The control Hamiltonian is
\begin{equation}
\label{hamiltonian}
H_c(p_1,p_2,x_1,x_2,u)=p_1x_2+\frac{p_2}{x_1^3}-p_2x_1u.
\end{equation}
Substituting (\ref{hamiltonian}) into (\ref{adjoint}) gives
\begin{eqnarray}
\label{adjoint1}
\dot{p}_1 & = & (u+\frac{3}{x_1^4})p_2,\\
\label{adjoint2}
\dot{p}_2 & = & -p_1.
\end{eqnarray}

According to the maximum principle, point 2 above, the time-optimal control $u(t)$ maximizes the control Hamiltonian at each time. Note that $H_c$ is a linear function of the control variable $u$. Since $u$ is bounded, $-v_1\leq u\leq v_2$, the optimal control that maximizes $H_c$ is determined by the sign of the coefficient of $u$, which is $-p_2x_1$. But $x_1>0$, thus when $p_2\neq 0$, the optimal control in $(0,t_f)$ is given by
\begin{equation}
\label{bang-bang}
u(t)=\left\{\begin{array}{cl} -v_1, & p_2>0 \\v_2, & p_2<0\end{array}\right..
\end{equation}
When
\begin{equation}
\label{nobang}
p_2=0
\end{equation}
for some time interval, the maximum principle provides a priori no information about the optimal control in this interval, which in that case is called a singular control. In general, singular extremals can play some role in the control of quantum systems \cite{Wu09, Lapert10}. We show that this is not the case for our problem, i.e. that condition (\ref{nobang}) cannot hold for any time interval $[t_1,t_2]\subset(0,t_f)$. Suppose that $p_2(t)=0$ for $t\in[t_1,t_2]$, then from (\ref{adjoint2}) we have $p_1=-\dot{p}_2=0$ for $t\in[t_1,t_2]$. Thus $p_1(t)=p_2(t)=0$ for $t\in[t_1,t_2]$, in contradiction with the maximum principle that requires the vector $\mathbf{p}(t)=(p_1(t),p_2(t))$ to be nonzero. So $p_2$ can be zero only at specific moments (switching times). The optimal control has ``bang-bang" form (\ref{bang-bang}), where the controller changes from one boundary value to the other at the switching times.

\begin{figure}[t]
\centering
\includegraphics[scale=0.35]{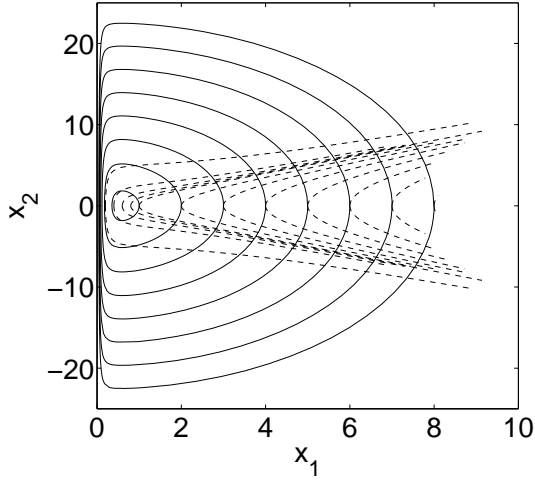}
\caption{Integral curves of the system  for $u=-v_1=-1$ (dashed line) and for $u=v_2=8$ (solid line).}
\label{fig:integral}
\end{figure}

Observe that when $u$ is a constant and Eqs. \eqref{system1} and \eqref{system2} are satisfied, then
\begin{equation}
\label{integraline}
x_2^2+ux_1^2+\frac{1}{x_1^2}=c,
\end{equation}
where $c$ is a constant. From \eqref{energy}, \eqref{definitions}  and \eqref{integraline} we find that $\langle H(t)\rangle_n/\hbar\omega_0=(2n+1)c/4$, so the paths of constant $u$ correspond to constant average energy for each mode.
In Fig. \ref{fig:integral} we plot the integral curves of the system defined in \eqref{system1} and \eqref{system2} for $u=-v_1$ and $u=v_2$.

\begin{figure}[t]
 \centering
		\begin{tabular}{cc}
     	\subfigure[$\ $Control function]{
	            \label{fig:one_s_con}
	            \includegraphics[width=.45\linewidth]{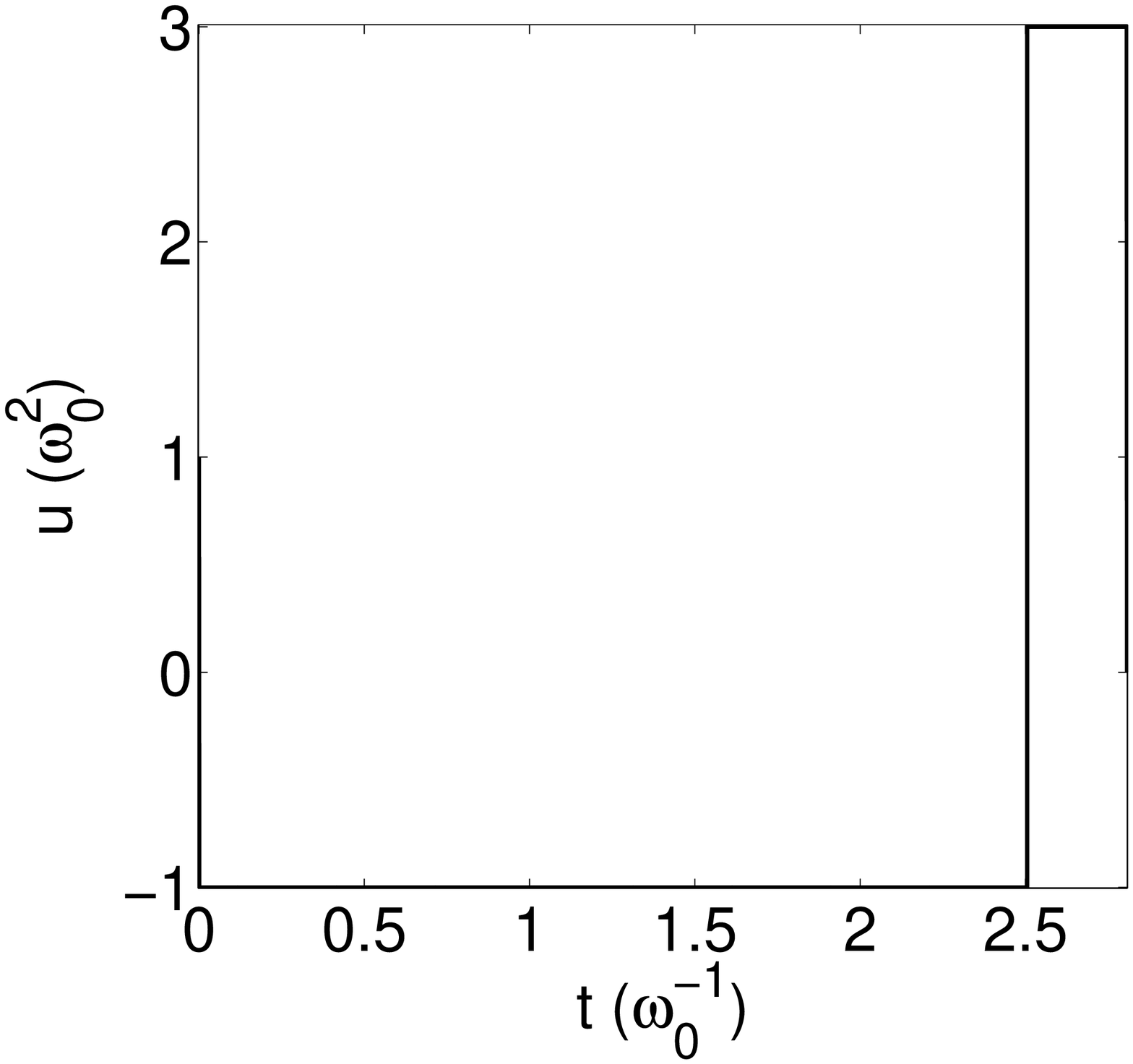}} &
	        \subfigure[$\ $Corresponding trajectory]{
	            \label{fig:one_s_traj}
	            \includegraphics[width=.45\linewidth]{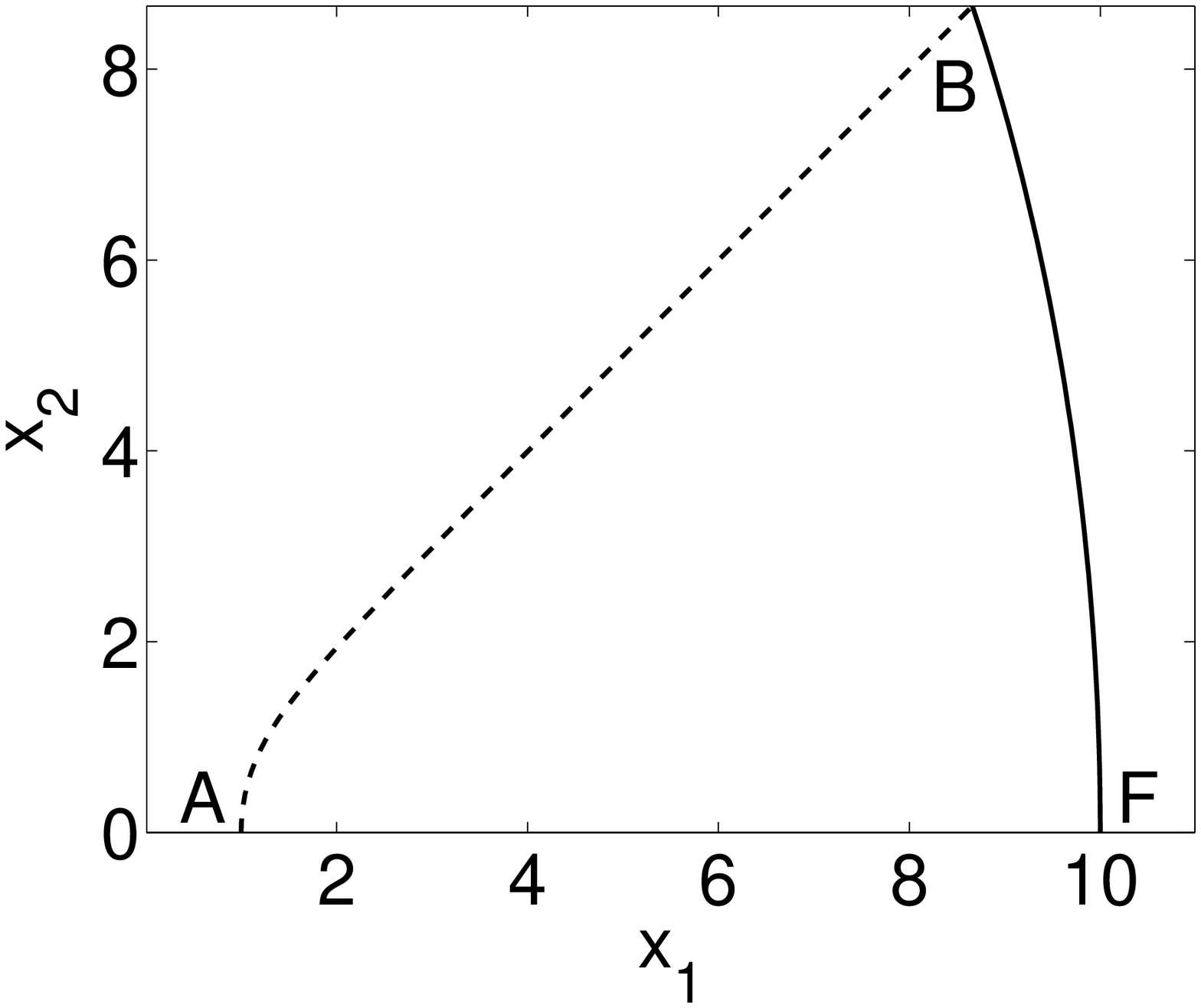}}
		\end{tabular}
\caption{The control function with one intermediate switching (panel a) and the corresponding trajectory (panel b) for $v_1=1,v_2=3$ and $\gamma=10$. Dashed line corresponds to $u=-v_1$, solid line to $u=v_2$.}
 \label{fig:oneswitching}
\end{figure}

For a feasible ``bang-bang" strategy with only one intermediate switching at $t=t_1$, the appropriate control sequence is
\begin{equation}
\label{oneswitching}
u(t)=\left\{\begin{array}{cl} 1, & t=0\\ -v_1, & 0<t<t_1\\ v_2, & t_1<t<t_1+t_2\\ 1/\gamma^4, & t=t_f^{(1)}=t_1+t_2\end{array}\right.,
\end{equation}
which is illustrated in Fig. \ref{fig:one_s_con}.
Applying the control boundary values in the opposite order does not transfer the state-space vector to the target. Note that the discontinuities at the beginning and at the end of the pulse sequence are not implied by the maximum principle but from the initial and final conditions on the control $u(t)$. If we ignore these boundary conditions and solve the corresponding time-optimal problem, the minimum time obtained is a lower bound of the minimum time when these conditions are on. This bound is achieved with instantaneous jumps of the control at the initial and final points.

We next calculate the necessary time to reach the final point following the control strategy (\ref{oneswitching}). Integrating \eqref{system1} and \eqref{system2} yields for $t\in[0,t_1]$
\begin{equation}
\label{x1phase1}
x_1(t)=\sqrt{1+\frac{v_1+1}{v_1}\sinh^2(\sqrt{v_1}t)},
\end{equation}
while for
$t\in[t_1,t_1+t_2]$
\begin{equation}
\label{x1phase2}
x_1(t)=\sqrt{\gamma^2-\frac{\gamma^4v_2-1}{\gamma^2v_2}\sin^2[\sqrt{v_2}(t_1+t_2-t)]}.
\end{equation}
From \eqref{integraline} we find that the state-space equation of the first segment $AB$ in Fig. \ref{fig:one_s_traj} is
\begin{equation}
\label{phase1}
x_2^2-v_1x_1^2+\frac{1}{x_1^2}=1-v_1,
\end{equation}
since $u=-v_1$ and the starting point $A(1,0)$ belongs to this segment. The corresponding equation for the second segment $BF$ is
\begin{equation}
\label{phase2}
x_2^2+v_2x_1^2+\frac{1}{x_1^2}=\gamma^2v_2+\frac{1}{\gamma^2},
\end{equation}
since $u=v_2$ and the final point $F(\gamma,0)$ belongs to this segment. The two segments meet at point $B(x_1^B,x_2^B)$. Subtracting (\ref{phase1}) from (\ref{phase2}) we find that
\begin{equation}
x_1^B=\sqrt{\frac{\gamma^2v_1+1+\gamma^2(\gamma^2v_2-1)}{\gamma^2(v_1+v_2)}}.
\end{equation}
Using the above value in (\ref{x1phase1}) and (\ref{x1phase2}) we obtain
\begin{eqnarray}
\label{time1}
t_1 & = &\frac{1}{\sqrt{v_1}}\sinh^{-1}\sqrt{\frac{v_1(\gamma^2-1)(\gamma^2v_2-1)}{\gamma^2(v_1+v_2)(v_1+1)}},\\
\label{time2}
t_2 & = &\frac{1}{\sqrt{v_2}}\sin^{-1}\sqrt{\frac{v_2(\gamma^2-1)(\gamma^2v_1+1)}{(v_1+v_2)(\gamma^4v_2-1)}}.
\end{eqnarray}
The total necessary time is
\begin{equation}
\label{timeoneswitching}
t_f^{(1)}=t_1+t_2,
\end{equation}
where the superscript denotes the number of intermediate switchings.

\begin{figure}[t]
 \centering
		\begin{tabular}{cc}
     	\subfigure[$\ $Intuitive control]{
	            \label{fig:two_s_con}
	            \includegraphics[width=.45\linewidth]{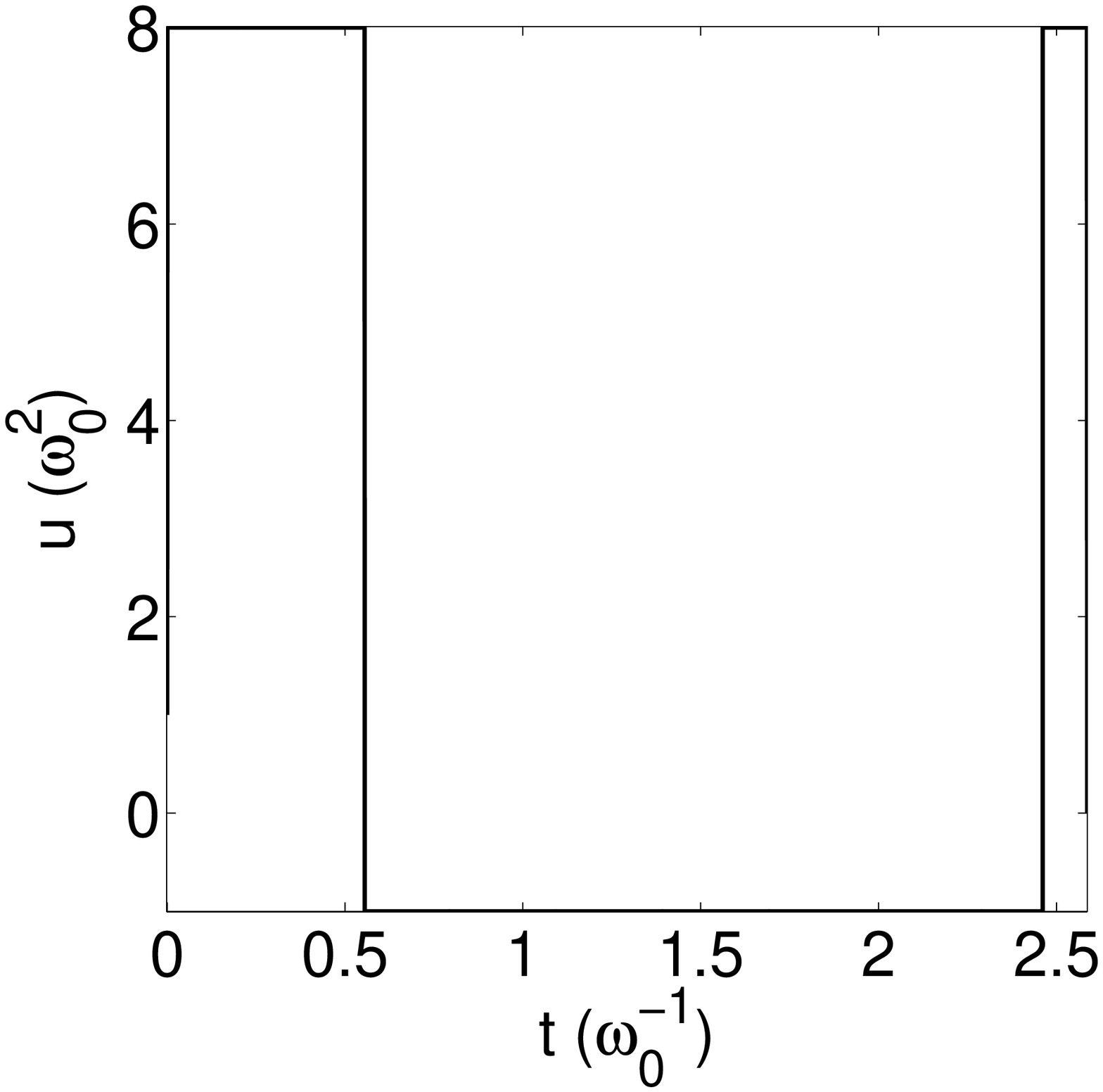}} &
	        \subfigure[$\ $Corresponding trajectory]{
	            \label{fig:twoswitchings}
	            \includegraphics[width=.45\linewidth]{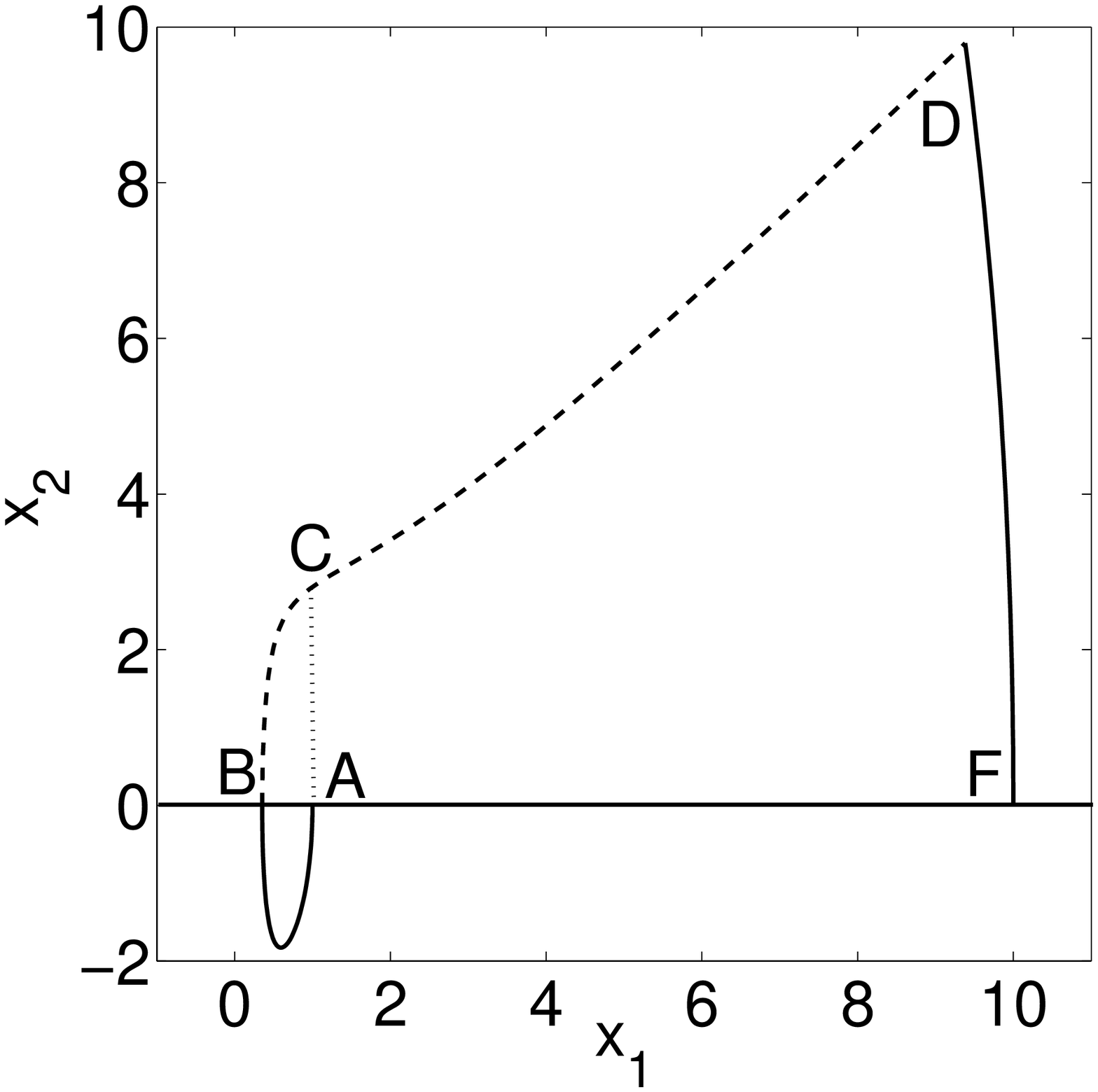}} \\
	        \subfigure[$\ $Optimal control]{
	            \label{fig:two_s_con1}
	            \includegraphics[width=.45\linewidth]{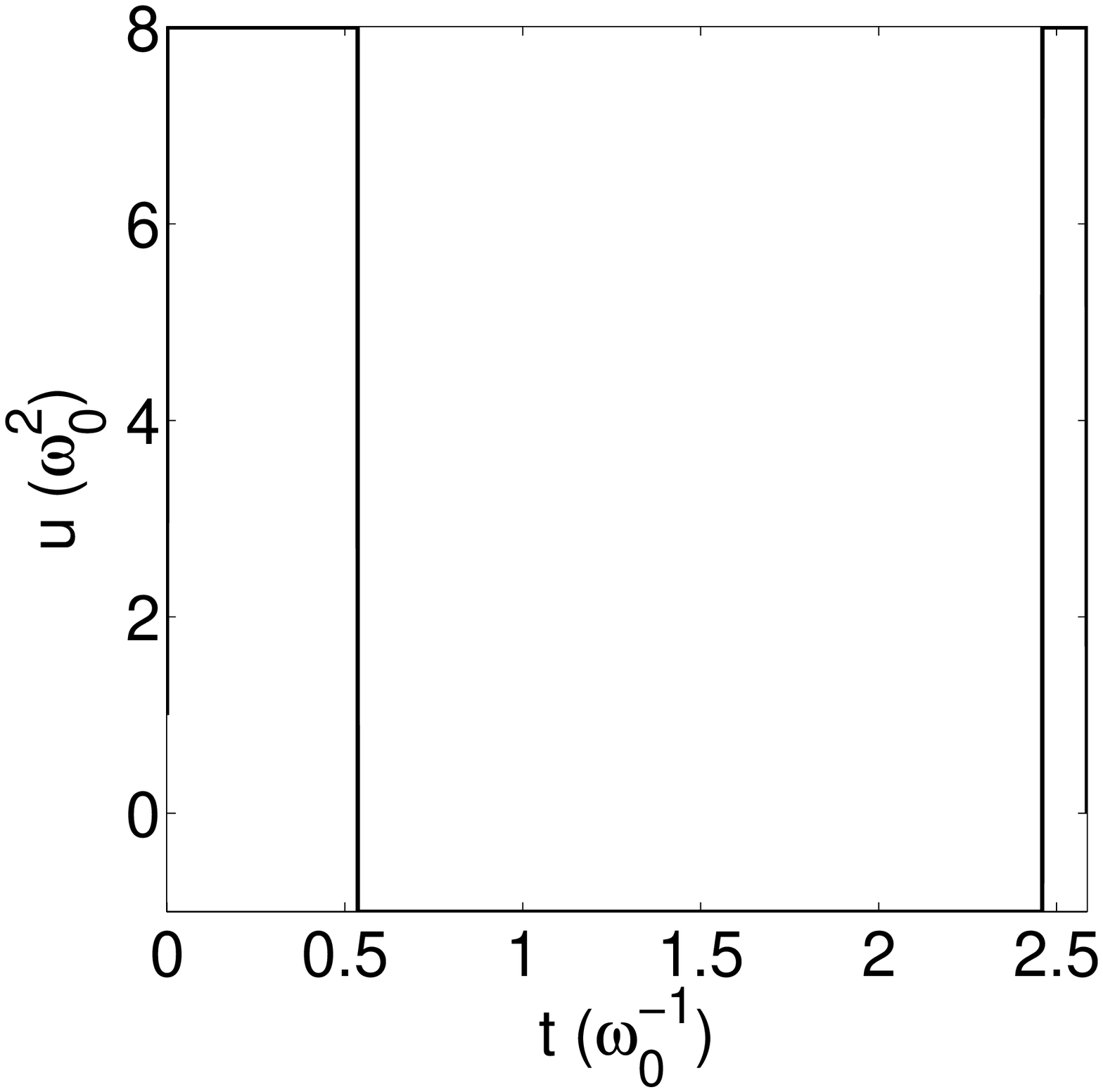}} &
			\subfigure[$\ $Corresponding trajectory]{
	            \label{fig:twoswitchings1}
	            \includegraphics[width=.45\linewidth]{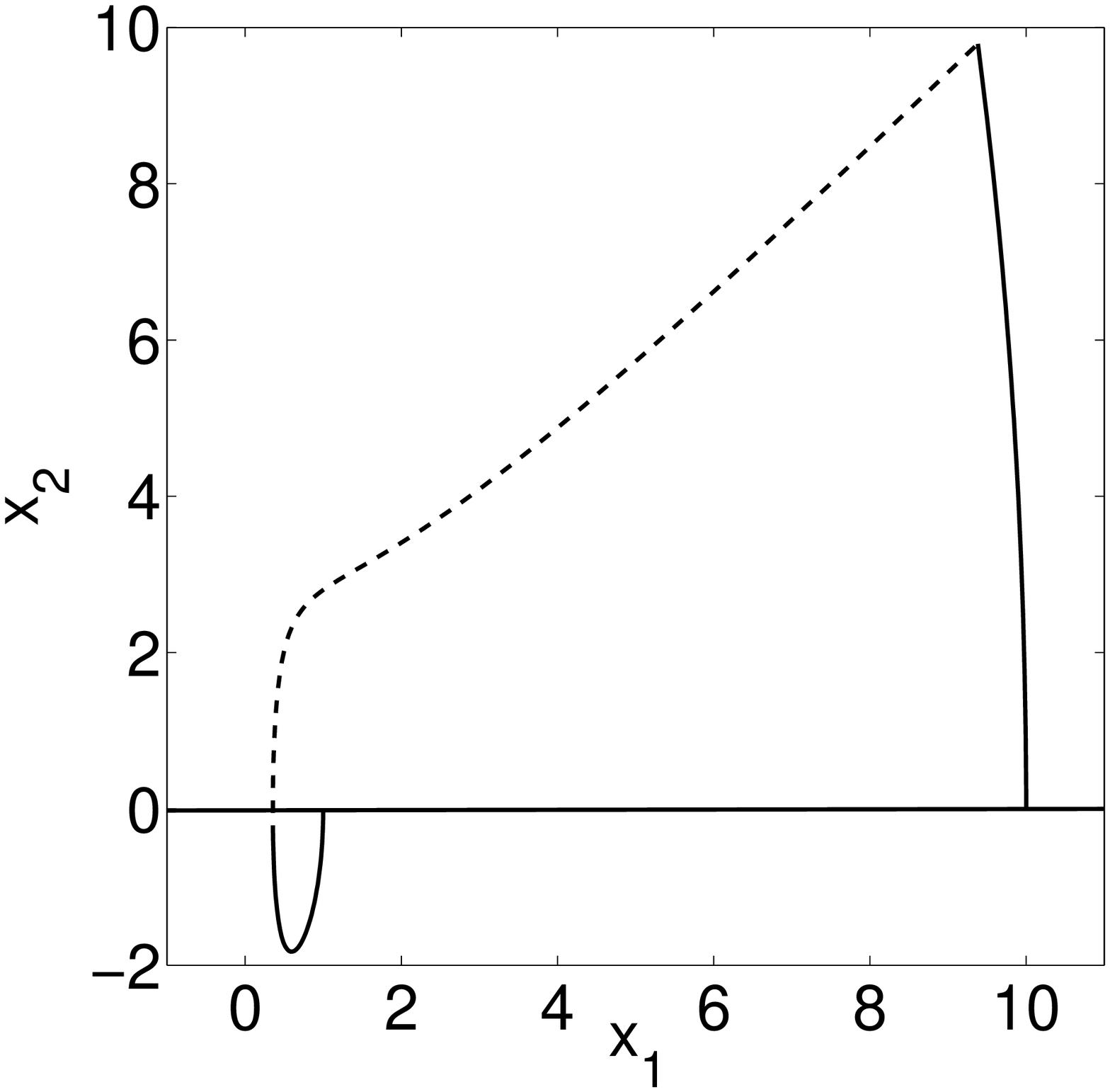}} \\
		\end{tabular}
\caption{The control functions with two intermediate switchings and the corresponding trajectories for $v_1=1,v_2=8,\gamma=10$. Dashed line corresponds to $u=-v_1$, solid line to $u=v_2$. Panels (a,b) show the intuitive solution with switching times given by (\ref{t1})-(\ref{t3}), while panels (c,d) show the optimal solution with at most two intermediate switchings where the switching times are calculated numerically.}
 \label{fig:switchingstwo}
\end{figure}

\begin{figure}[t]
\centering
		\begin{tabular}{cc}
    	\subfigure[$\ $Transfer times for one and two intermediate switchings]{
	            \label{fig:difference}
	            \includegraphics[width=0.45\linewidth]{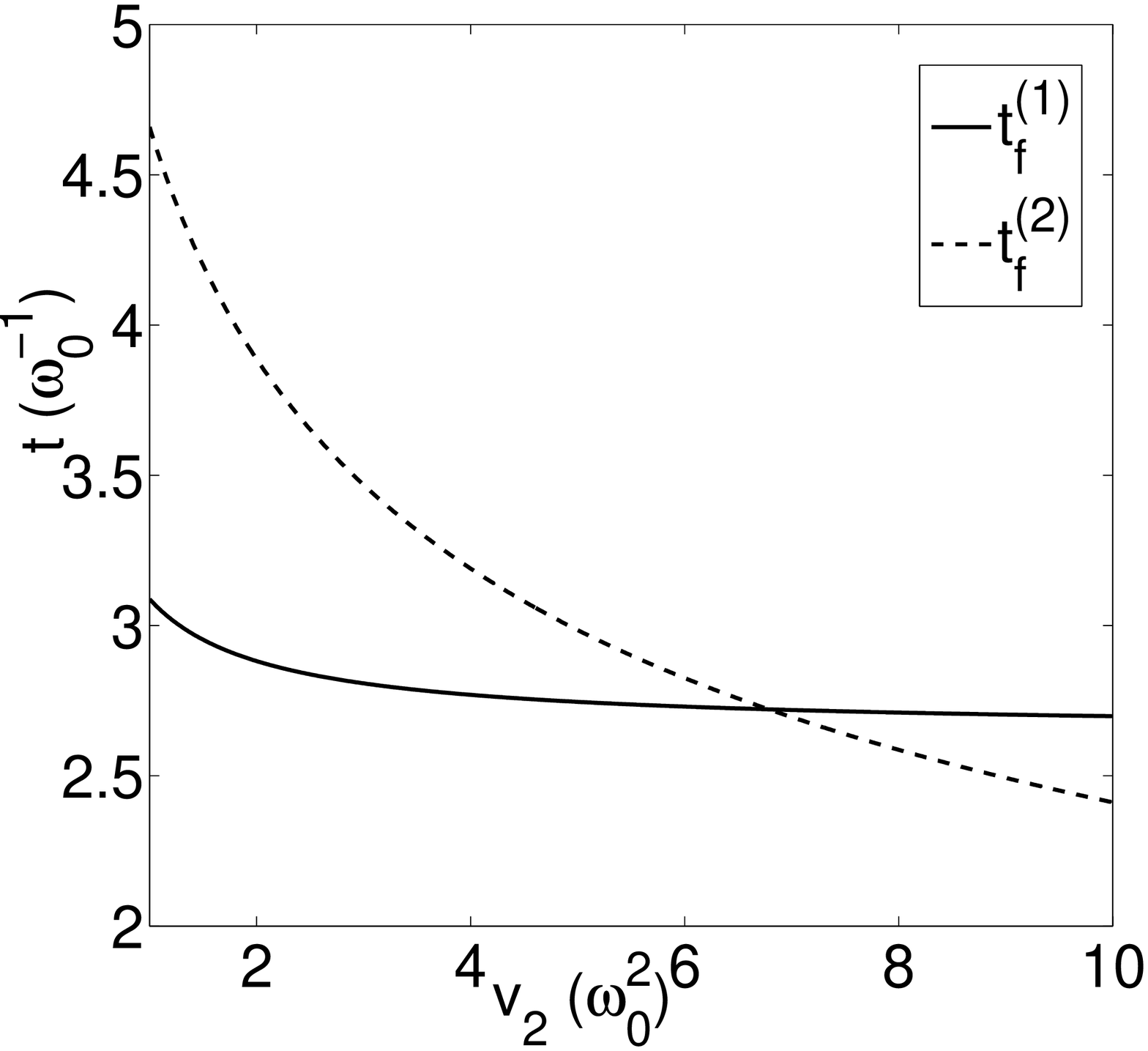}} &
	        \subfigure[$\ $Optimal transfer time for at most two intermediate switchings]{
	            \label{fig:difference1}
	            \includegraphics[width=0.45\linewidth]{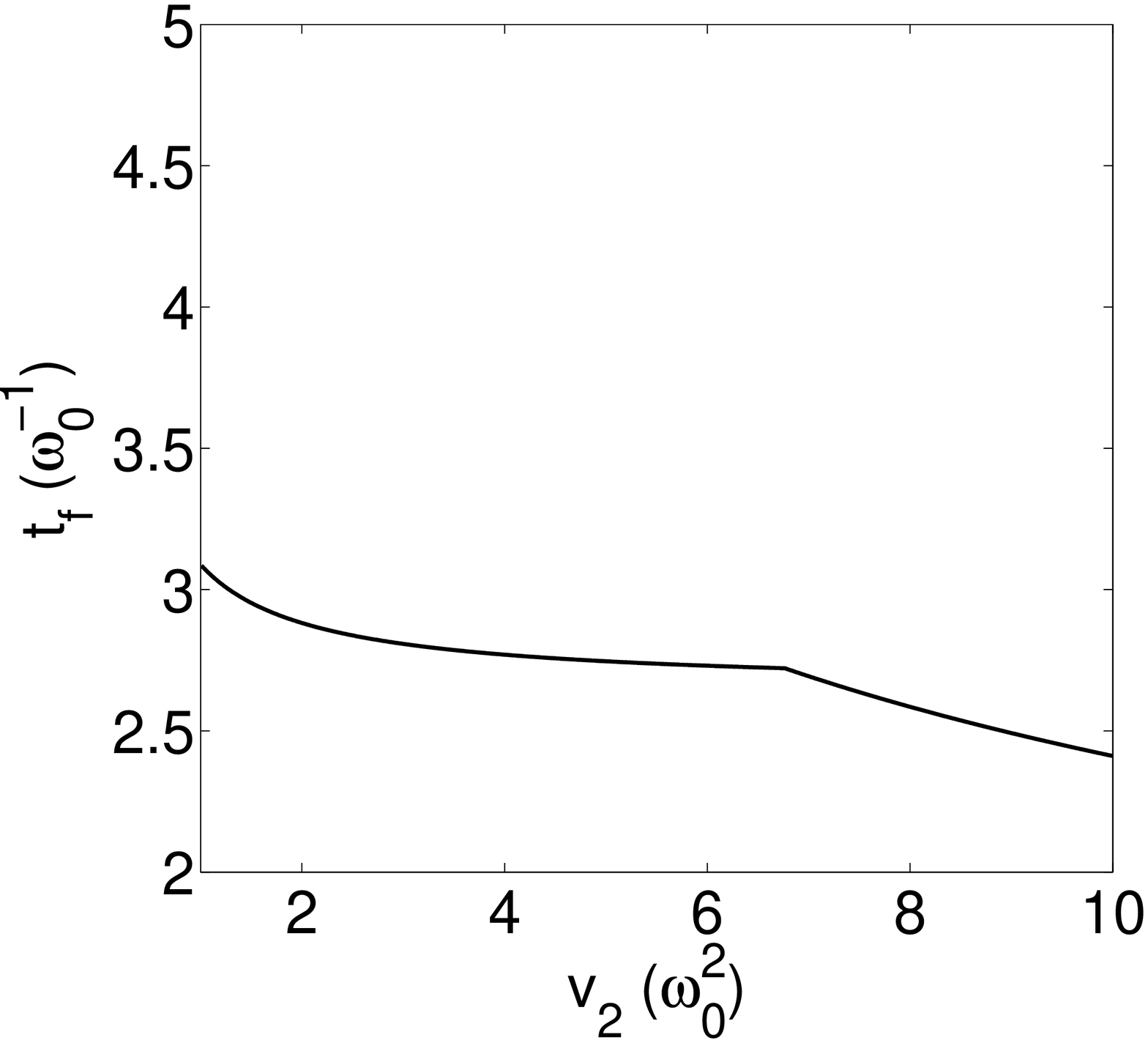}}
		\end{tabular}
\caption{Total transfer times for the presented strategies as a function of $v_2$ for $v_1=1$ and $\gamma=10$. In panel (a) we plot the transfer time $t_f^{(1)}$ of the strategy with one intermediate switching (solid line), as well as the transfer time $t_f^{(2)}$ of the intuitive solution with two intermediate switchings (dashed line). Observe that for $v_2>6.786$ the second strategy is faster. In panel (b) we plot the transfer time $t_f$ for the optimal strategy with at most two intermediate switchings, retrieving similar results.}
\end{figure}

We next show that when $v_2$ is large enough we can find a control strategy with two intermediate switchings that accomplishes the desired transfer in less time. Consider the following control sequence
\begin{equation}
\label{twoswitchings}
u(t)=\left\{\begin{array}{cl} 1, & t=0\\ v_2, & 0<t<t_1\\-v_1, & t_1<t<t_1+t_2\\v_2, & t_1+t_2<t<t_1+t_2+t_3\\ 1/\gamma^4, & t=t_f^{(2)}=t_1+t_2+t_3\end{array}\right.
\end{equation}
where
\begin{eqnarray}
\label{t1}
t_1 & = & \frac{1}{2}\frac{\pi}{\sqrt{v_2}},\\
\label{t2}
t_2 & = &\frac{1}{\sqrt{v_1}}\sinh^{-1}\sqrt{\frac{v_1v_2(\gamma^2-1)(\gamma^2v_2-1)}{\gamma^2(v_1+v_2)(v_1+v_2^2)}},\\
\label{t3}
t_3 & = &\frac{1}{\sqrt{v_2}}\sin^{-1}{\sqrt{\frac{(\gamma^2v_2-1)(v_2+\gamma^2v_1)}{(v_1+v_2)(\gamma^4v_2-1)}}}.
\end{eqnarray}
Time $t_1$ is chosen such that the first intermediate switching takes place as close as possible to $x_1=0$ (we explain later how this is related to minimizing the transfer time), while $t_2$ and $t_3$ are determined such that the target point $F(\gamma,0)$ is reached at the final time.
The control $u(t)$ and the corresponding trajectory for $v_1=1,v_2=8$ and $\gamma=10$ are shown in Figs. \ref{fig:two_s_con} and \ref{fig:twoswitchings}, respectively.
The total necessary time for this control policy is
\begin{equation}
\label{timetwoswitchings}
t_f^{(2)}=t_1+t_2+t_3.
\end{equation}
Observe that for $v_1$ constant
\begin{eqnarray}
\mbox{lim}_{v_2\rightarrow\infty}\,t_f^{(1)} & = & \frac{1}{\sqrt{v_1}}\sinh^{-1}\sqrt{\frac{v_1(\gamma^2-1)}{v_1+1}},\\
\mbox{lim}_{v_2\rightarrow\infty}\,t_f^{(2)} & = & 0,
\end{eqnarray}
thus there is a value $v_2^*$ such that $t_f^{(2)}<t_f^{(1)}$ for $v_2>v_2^*$. In Fig. \ref{fig:difference} we plot $t_f^{(1)}$ and $t_f^{(2)}$ as a function of $v_2$ for $v_1=1$ and $\gamma=10$. Observe that for $v_2>6.786$, the strategy including two intermediate switchings is faster.

To understand intuitively why such a strategy can transfer the state vector to the final point in less time, we use the one-dimensional particle model where the position $x_1$ and velocity $x_2$ of a unit mass particle satisfy equations \eqref{system1} and \eqref{system2}, and refer to Fig. \ref{fig:twoswitchings}. If $v_2$ is large enough, then the particle can be transferred relatively fast from starting point $A$, with position $x_1=1$ and velocity $x_2=0$, to point $B$, with $0<x_1\ll 1$. At this point, the force term $1/x_1^3$ is very large and substantially accelerates the particle. When the particle passes through point $C$, with position $x_1=1$ same as the starting point $A$, it now has a nonzero velocity ($x_2>0$) that allows it to travel faster at the final point $F$, with $x_1=\gamma$. The repulsive potential at $x_1=0$ acts as a slingshot, resembling the gravitational slingshots used in aerospace engineering to alter the speed of a spacecraft.

We emphasize that the values of $t_1,t_2$ and $t_3$ given by \eqref{t1}-\eqref{t3} are not optimal but correspond to a suboptimal intuitive solution.
We can determine the optimal switching times numerically if we vary $t_1$ (for specific $t_1$, $t_2$ and $t_3$ are automatically fixed from the requirement to reach the target point at the final time) and pick the value that minimizes the total transfer time. In Figs. \ref{fig:two_s_con1} and \ref{fig:twoswitchings1} we show the numerically calculated optimal solution with two intermediate switchings for the same parameter values that are used in Figs. \ref{fig:two_s_con} and \ref{fig:twoswitchings} for the intuitive solution. Observe that the two solutions are very similar but in the optimal one the first intermediate jump takes place slightly earlier, before the $x_1$ axis is reached. In Fig. \ref{fig:difference1} we plot the total transfer time $t_f$, obtained with this numerical method, as a function of $v_2$ for $v_1=1$ and $\gamma=10$. Comparing with Fig. \ref{fig:difference} it is not hard to find that for $v_2<6.763$ it is $t_f=t_f^{(1)}$, while for $v_2>6.763$ it is $t_f\approx t_f^{(2)}$ (actually $t_f<t_f^{(2)}$ as expected since the intuitive solution is suboptimal). In other words, for $v_2<6.763$ the numerically calculated optimal solution has only one intermediate switching (the method gives $t_1=0$ indeed), while for $v_2>6.763$ is very close to the intuitive solution given by (\ref{twoswitchings})-(\ref{t3}). Note that since $t_f<t_f^{(2)}$, the transition value of $v_2$ just found (6.763) is slightly lower than the one found above (6.786) using the intuitive solution.

\begin{figure}[t]
 \centering
		\begin{tabular}{cc}
     	\subfigure[$\ $Trajectory with $2n$ switchings]{
	            \label{fig:twon}
	            \includegraphics[width=.45\linewidth]{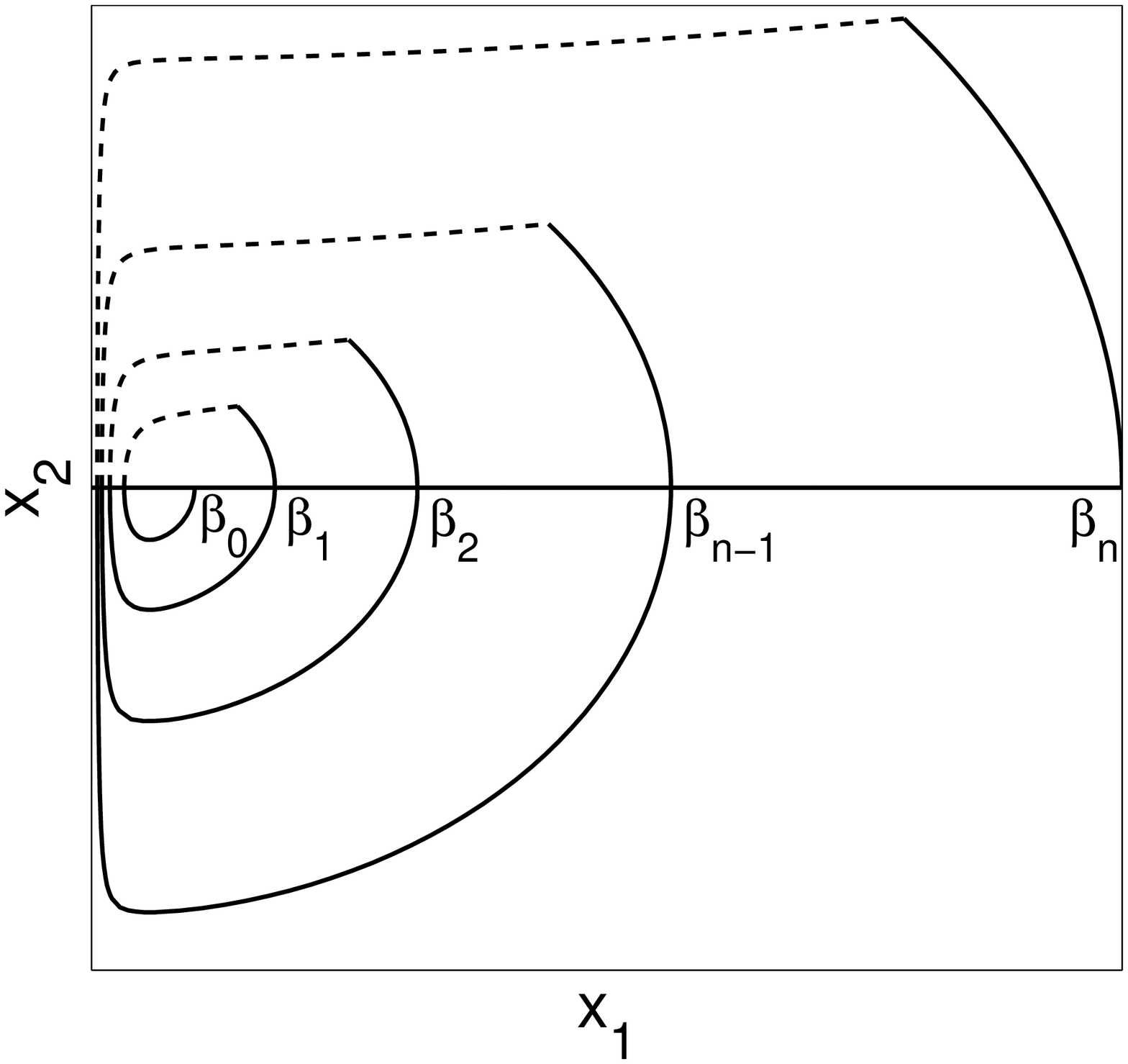}} &
	        \subfigure[$\ $Segment with two switchings]{
	            \label{fig:two}
	            \includegraphics[width=.45\linewidth]{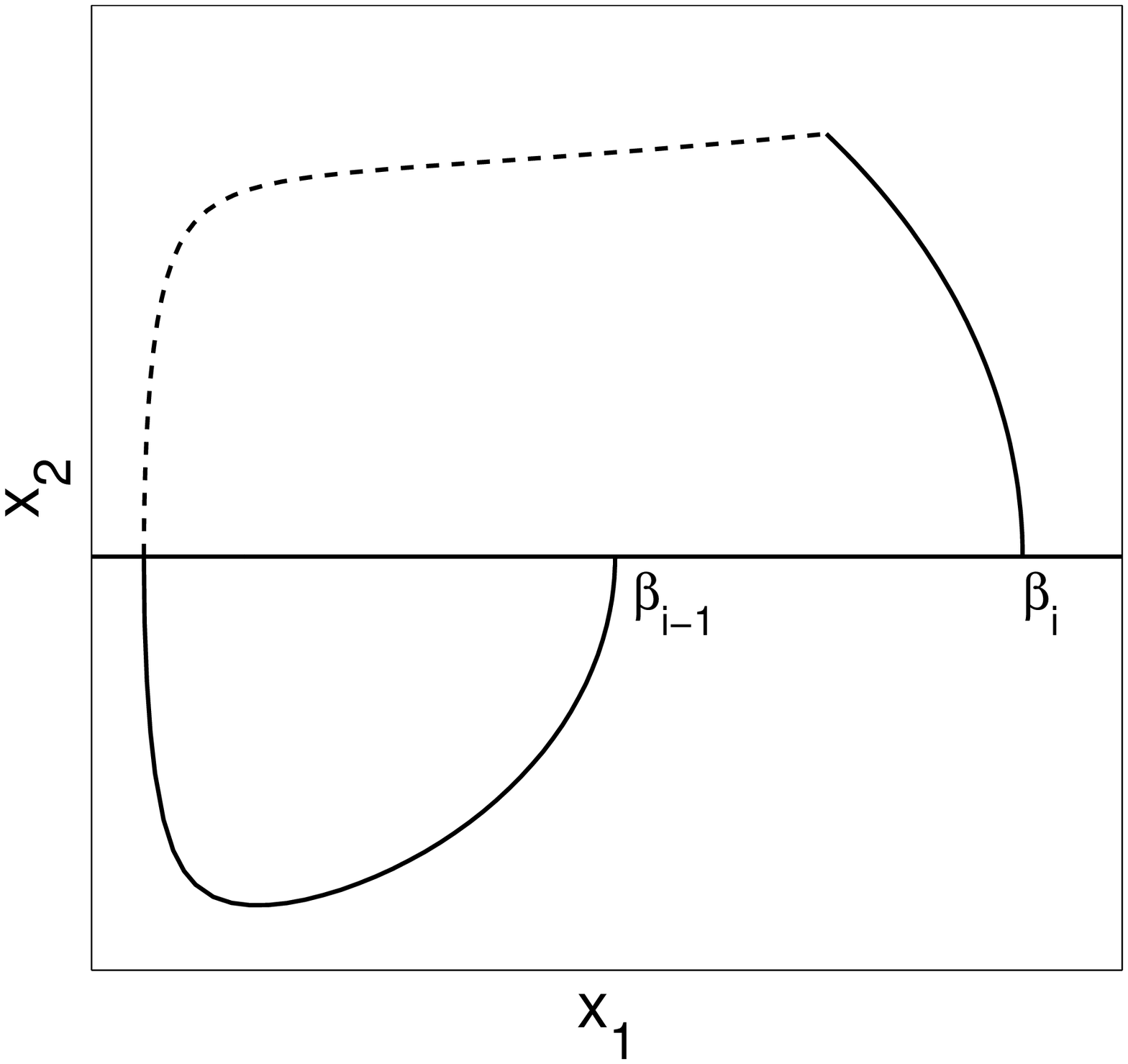}} \\
	        \subfigure[$\ $Transfer time $t^{(2n)}_f$ as $v_2\rightarrow\infty$ for optimal $\beta_i$]{
	            \label{fig:time2n}
	            \includegraphics[width=.45\linewidth]{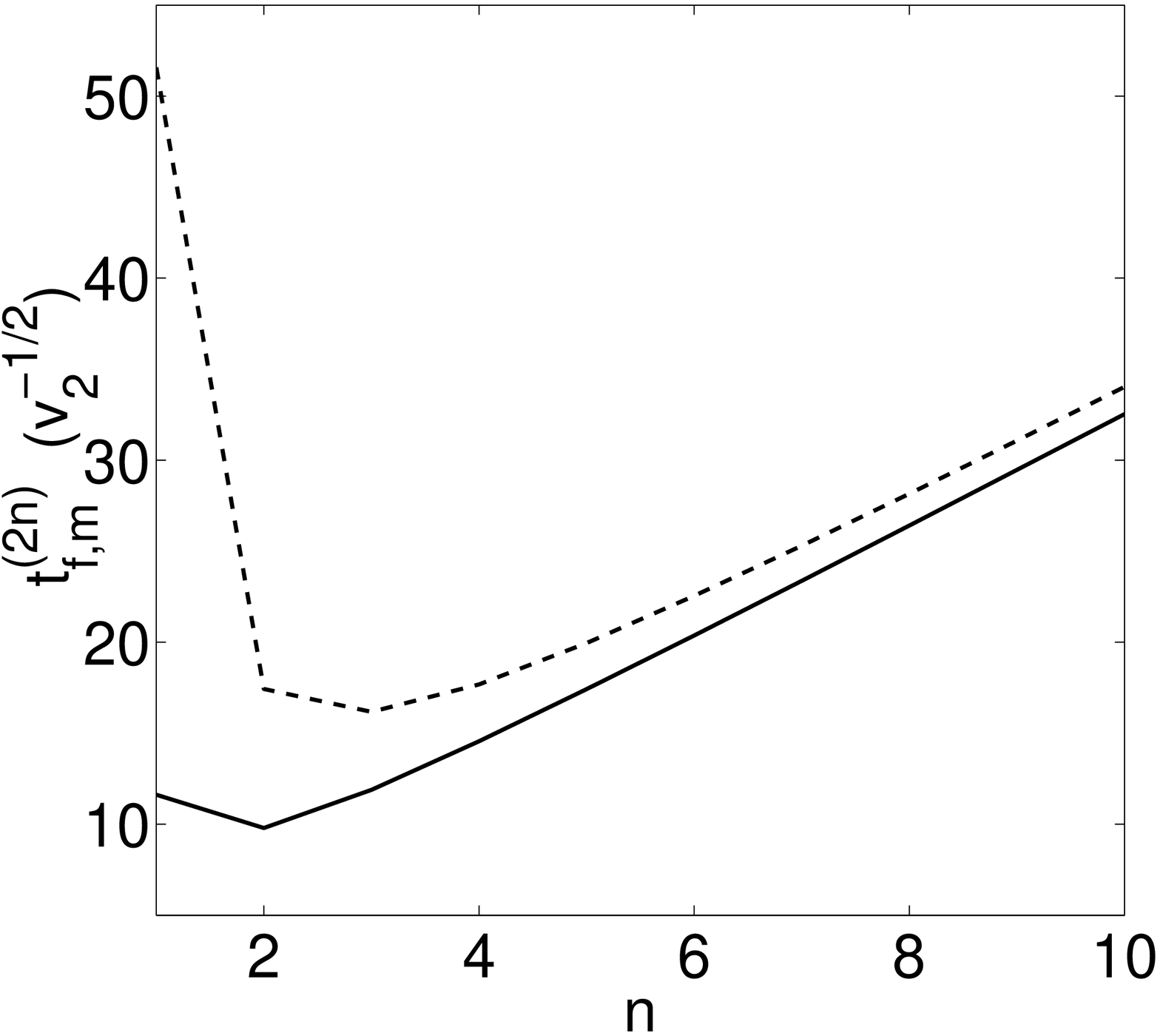}} &
			\subfigure[$\ $Exact transfer times for two, four and six switchings]{
	            \label{fig:time246}
	            \includegraphics[width=.45\linewidth]{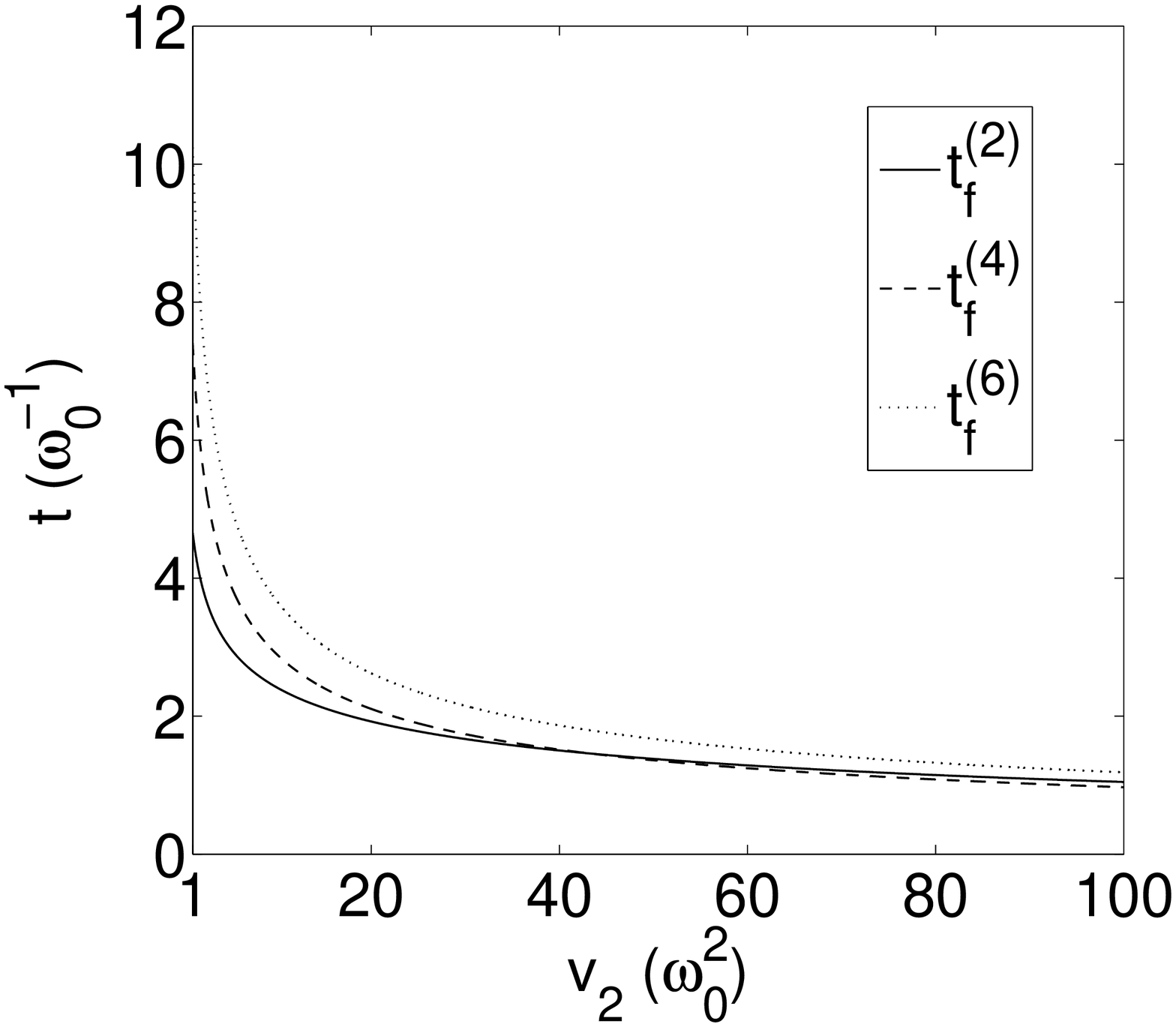}} \\
		\end{tabular}
\caption{A trajectory with $2n$ intermediate switchings (panel a) composed by $n$ segments with two switchings (panel b). Dashed line corresponds to $u=-v_1$, solid line to $u=v_2$. Note that $\beta_0=1$ and $\beta_n=\gamma$. In the limit $v_2\rightarrow\infty$ the minimum transfer time corresponding to the optimal choice of $\beta_i$ is plotted (panel c) for $\gamma=10$ (solid line) and $\gamma=50$ (dashed line) as a function of $n$. Observe that for larger values of $\gamma$ the minimum is shifted towards larger values of $n$. For $\gamma=10$ it is achieved for $n=2$, i.e. four intermediate switchings. The exact transfer times $t^{(2)}_f,t^{(4)}_f$ and $t^{(6)}_f$, as calculated from \eqref{two_segment} (valid at all scales of $v_2$) are plotted as functions of $v_2$ for $\gamma=10$ and $v_1=1$. Observe that for $v_2\geq 43.32$ the four switchings strategy (dashed line) becomes indeed optimal among the control policies that we consider.}
 \label{fig:2n}
\end{figure}
For larger values of $v_2$ it is possible to find strategies with more than two intermediate switchings with shorter transfer times, depending on the value of the target coordinate $\gamma$. For example, consider the strategy with $2n$ intermediate switchings whose corresponding trajectory is shown in Fig \ref{fig:twon}. It is actually composed by $n$ segments with two switchings, see Fig. \ref{fig:two}. The necessary time to travel the segment starting from $(\beta_{i-1},0)$ and ending at $(\beta_i,0)$ is
\begin{multline}
\label{two_segment}
t^{(2)}_f(\beta_{i-1},\beta_i)=t_1\left(\frac{1}{\beta_{i-1}\sqrt{v_2}},\beta_i\right)+t_2\left(\frac{1}{\beta_{i-1}\sqrt{v_2}},\beta_i\right)+\\\frac{\pi}{2\sqrt{v_2}},
\end{multline}
where
\begin{eqnarray}
\label{alpha_beta_1}
t_1(\alpha,\beta) & = & \frac{1}{\sqrt{v_1}}\sinh^{-1}\sqrt{\frac{v_1(\beta^2-\alpha^2)(\alpha^2\beta^2v_2-1)}{\beta^2(v_1+v_2)(\alpha^4v_1+1)}},\nonumber\\
\label{alpha_beta_2}
t_2(\alpha,\beta) & = & \frac{1}{\sqrt{v_2}}\sin^{-1}\sqrt{\frac{v_2(\beta^2-\alpha^2)(\alpha^2\beta^2v_1+1)}{\alpha^2(v_1+v_2)(\beta^4v_2-1)}}.\nonumber
\end{eqnarray}
In the limit $v_2\rightarrow\infty$ we obtain
\begin{equation}
\label{two_segment_limit}
t^{(2)}_f(\beta_{i-1},\beta_i)\rightarrow\frac{1}{\sqrt{v_2}}\left[\frac{\pi}{2}+\sqrt{\frac{\beta_i^2}{\beta_{i-1}^2}-1}+\sin^{-1}\left(\frac{\beta_{i-1}}{\beta_i}\right)\right]
\end{equation}
The total transfer time for the strategy with $2n$ switchings is
\begin{equation}
\label{2n}
t^{(2n)}_f=\sum_{i=1}^nt^{(2)}_f(\beta_{i-1},\beta_i),
\end{equation}
where $\beta_0=1$ and $\beta_n=\gamma$. We can find the optimal $\beta_i,i=1,2,\ldots n-1$ that minimize $t^{(2n)}_f$ using dynamic programming. Suppose that we know the optimal $\beta_{i}, i=1,2,\ldots n-2$ and we want to find $\beta_{n-1}$. This variable appears only in the terms $t^{(2)}_f(\beta_{n-2},\beta_{n-1})$ and $t^{(2)}_f(\beta_{n-1},\beta_n)$ of the sum \eqref{2n}. Using Eq. \eqref{two_segment_limit} to approximate these terms and equating with zero the derivative of their sum with respect to this variable, we find that the optimal $\beta_{n-1}$ satisfies $\beta^2_{n-1}=\beta_{n-2}\beta_n$ in the limit $v_2\rightarrow\infty$. It corresponds to a minimum since the second derivative can be easily found to be positive. Working analogously we get
\begin{equation}
\label{geometric_mean}
\beta^2_{i}=\beta_{i-1}\beta_{i+1}
\end{equation}
for $i=1,2,\ldots n-1$.
Since $\beta_0=1,\beta_n=\gamma$, we obtain $\beta_i=\gamma^{i/n}, i=1,2,\ldots n$. The minimum value $t^{(2n)}_{f,m}$ of the transfer time $t^{(2n)}_f$, as $v_2\rightarrow\infty$, is
\begin{equation}
\label{2n_min_time}
t^{(2n)}_{f,m}=\frac{n}{\sqrt{v_2}}\left[\frac{\pi}{2}+\sqrt{\gamma^{2/n}-1}+\sin^{-1}\left(\frac{1}{\gamma^{1/n}}\right)\right],
\end{equation}
where note that each of the $n$ segments is traveled in equal time.
In Fig \ref{fig:time2n} we plot $t^{(2n)}_{f,m}$ in units of $1/\sqrt{v_2}$ as a function of $n$ for $\gamma=10$ (solid line) and $\gamma=50$ (dashed line). Observe that for larger $\gamma$ the minimum of $t^{(2n)}_{f,m}$ is shifted towards larger values of $n$, i.e. the particle passes more times close to $x_1=0$ to acquire more speed and thus reach faster the more distant target point. For $\gamma=10$ the minimum is obtained for $n=2$, i.e. a four switchings strategy. In Fig \ref{fig:time246} we plot $t^{(2n)}_f, n=1,2,3$, for $\gamma=10$ and $v_1=1$, using the exact formula \eqref{two_segment} and not the approximation \eqref{two_segment_limit}. Observe that for $v_2\geq43.32$ the four switchings strategy is indeed the optimal among the control policies that we considered.

Although the time-optimal control has ``bang-bang" form, as shown above, such discontinuous changes in $u(t)$ are unrealistic and difficult to implement experimentally. To overcome this problem, in the next section we use a powerful numerical optimization method that allows us to find realistic time-optimal solutions, following the path introduced in \cite{Chen10}.

\section{Realistic solutions using a pseudospectral numerical method}

Pseudospectral methods were developed to solve partial differential equations and recently adapted to solve optimal control problems, see for example \cite{elnagar_pseudospectral_1995, ross_legendre_2004, wei_kang_convergence_2006, fahroo_costate_2001, williams_gauss--lobatto_2006} and our recent work for optimal pulse design in Nuclear Magnetic Resonance spectroscopy \cite{Li09}. They are used to convert a continuous-time optimal control problem to a discrete nonlinear programming problem, which can be solved by many well developed computational algorithms.

These methods involve the approximation of the control and state functions, $u(t)$ and $\mathbf{x}(t)$, by orthogonal polynomial basis functions on the domain $[-1,1]$. Using such a basis leads to \emph{spectral accuracy}, namely, the $k^{\text{th}}$ coefficient of the expansion decays faster than any inverse power of $k$ \cite{canuto_spectral_2006}, permitting the use of relatively low order polynomials in the approximations.

\begin{figure}[t]
 \centering
		\begin{tabular}{cc}
     	\subfigure[$\ $Uniform Grid]{
	            \label{fig:uniform}
	            \includegraphics[width=.5\linewidth]{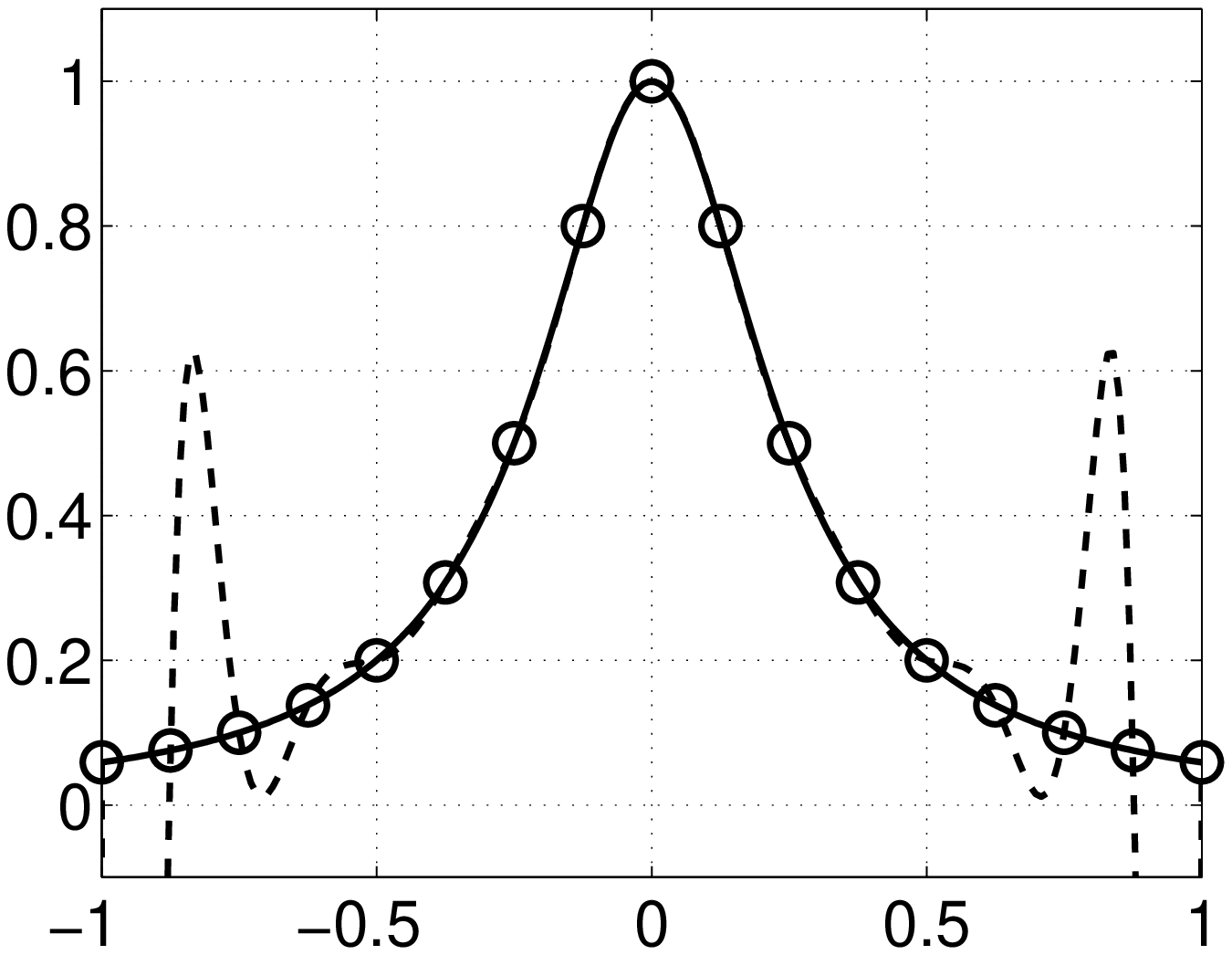}} &
	        \subfigure[$\ $LGL Grid]{
	            \label{fig:lgl}
	            \includegraphics[width=.5\linewidth]{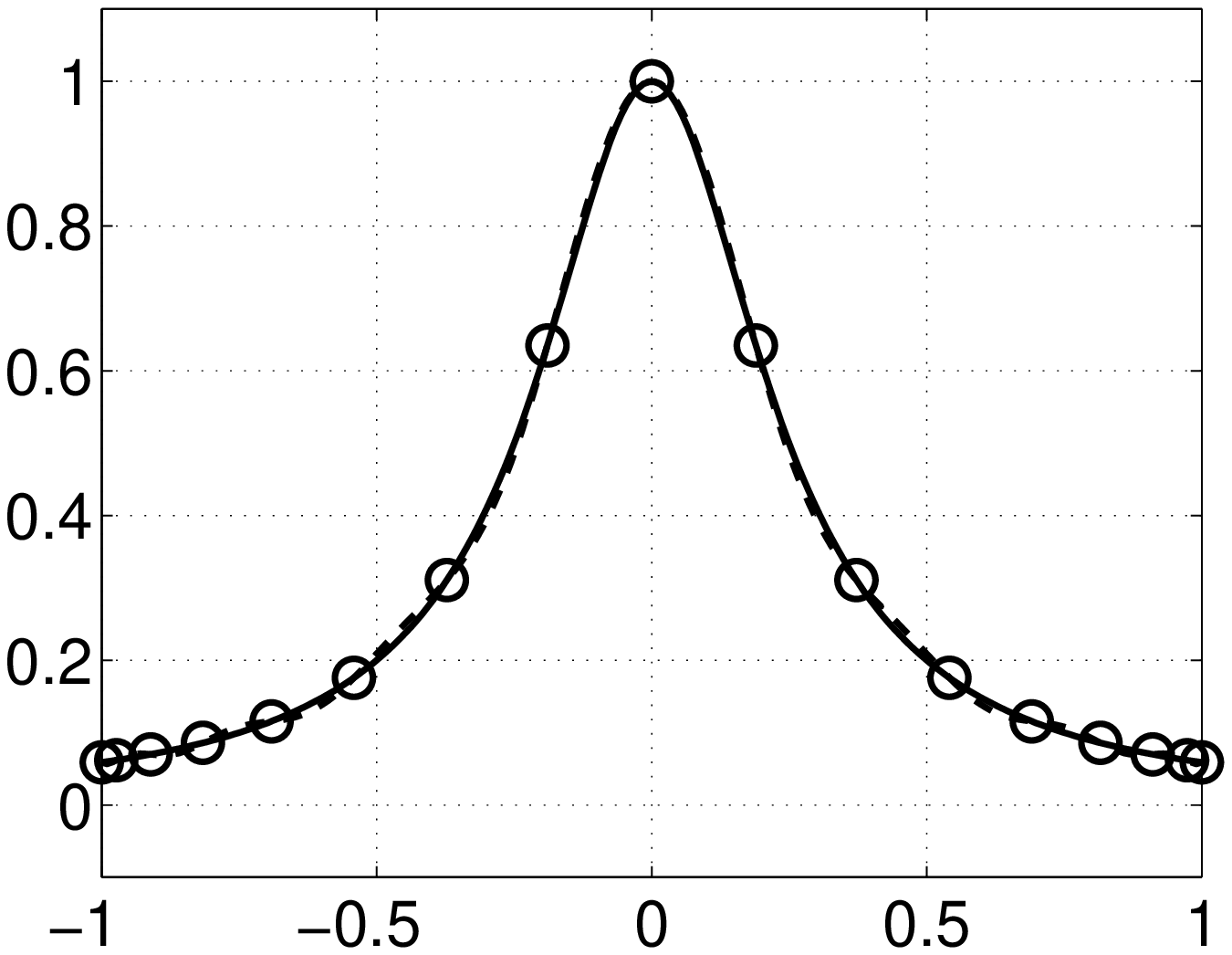}}
		\end{tabular}
 \caption{The $N=16$ order interpolation of the function $f(x)=1/(16x^2+1)$ based on a uniform grid (panel a) demonstrates the Runge phenomenon whereas the interpolation based on the LGL grid (panel b) does not.}
 \label{fig:uniform_lgl}
\end{figure}

In order to apply such a method, the first step is to transform the optimal control problem from the time domain $t\in[0,t_f]$ to $\tau\in[-1,1]$ using the simple affine transformation $\tau(t)=(2t-t_f)/t_f$. In a redundant use of notation, we make this transition and reuse the same time variable $t$. The system equations become
\begin{eqnarray}
\label{modsystem1}
\dot{x}_1 & = & \frac{t_f}{2}\,x_2\\
\label{modsystem2}
\dot{x}_2 & = & \frac{t_f}{2}\left(-ux_1+\frac{1}{x_1^3}\right)
\end{eqnarray}
with boundary conditions $(x_1(-1),x_2(-1),u(-1))=(1,0,1)$ and $(x_1(1),x_2(1),u(1))=(\gamma,0,1/\gamma^4)$.

According to the Chebyshev Equioscillation Theorem \cite{davis_interpolation_1963} the best $N^{\text{th}}$ order approximating polynomial to a continuous function on the interval $[-1,1]$, as evaluated by the uniform norm, is an interpolating polynomial. Since any $N^{\text{th}}$ order interpolating polynomial can be represented in terms of the Lagrange polynomials, we use these functions to express the interpolating approximations of the state and control functions, $\mathbf{x}(t)$ and $u(t)$. Given a grid of $N+1$ interpolation nodes within $[-1,1]$, $\Gamma = \{t_0 < t_1 < \dots < t_N\}$, the Lagrange polynomials $\{\ell_k\}$, $k\in\{0,1,\ldots,N\}$, are constructed by
\begin{equation*}
    \ell_k(t) = \prod_{i=0 \atop i\ne k}^N \displaystyle\frac{(t-t_i)}{(t_k-t_i)}.
\end{equation*}
They form an orthogonal basis with respect to the discrete inner product $\langle p,q \rangle =\sum_{k=0}^{N}p(t_k)q(t_k)$, while $\ell_k(t_i) = \delta_{ki}$ holds at the grid nodes \cite{szego_orthogonal_1959}.

\begin{figure}[t]
 \centering
		\begin{tabular}{cc}
     	\subfigure[$\ $Optimal control, $M=\infty$]{
	            \label{fig:ps_con_1}
	            \includegraphics[width=.45\linewidth]{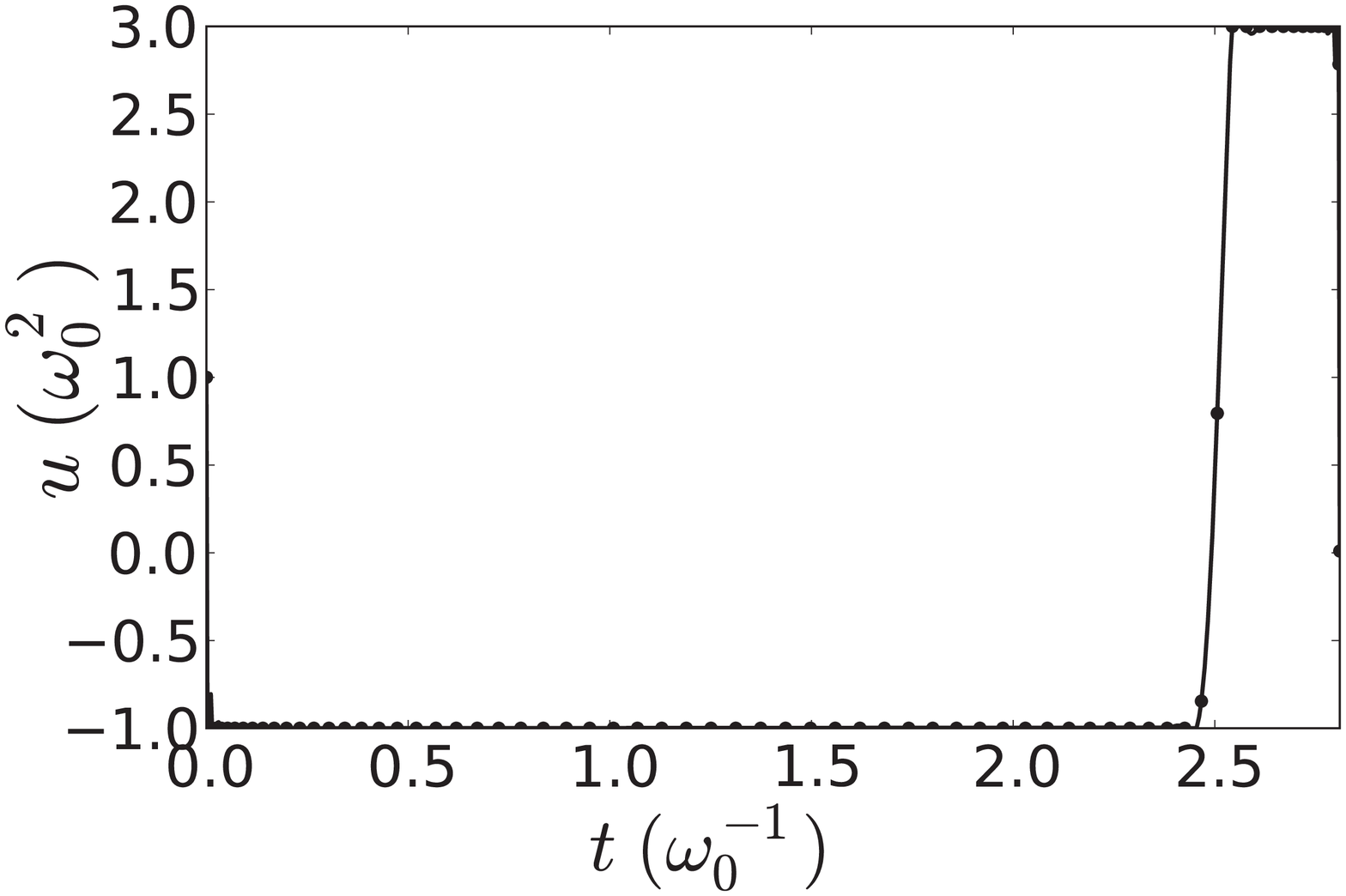}} &
	        \subfigure[$\ $Corresponding trajectory]{
	            \label{fig:ps_traj_1}
	            \includegraphics[width=.45\linewidth]{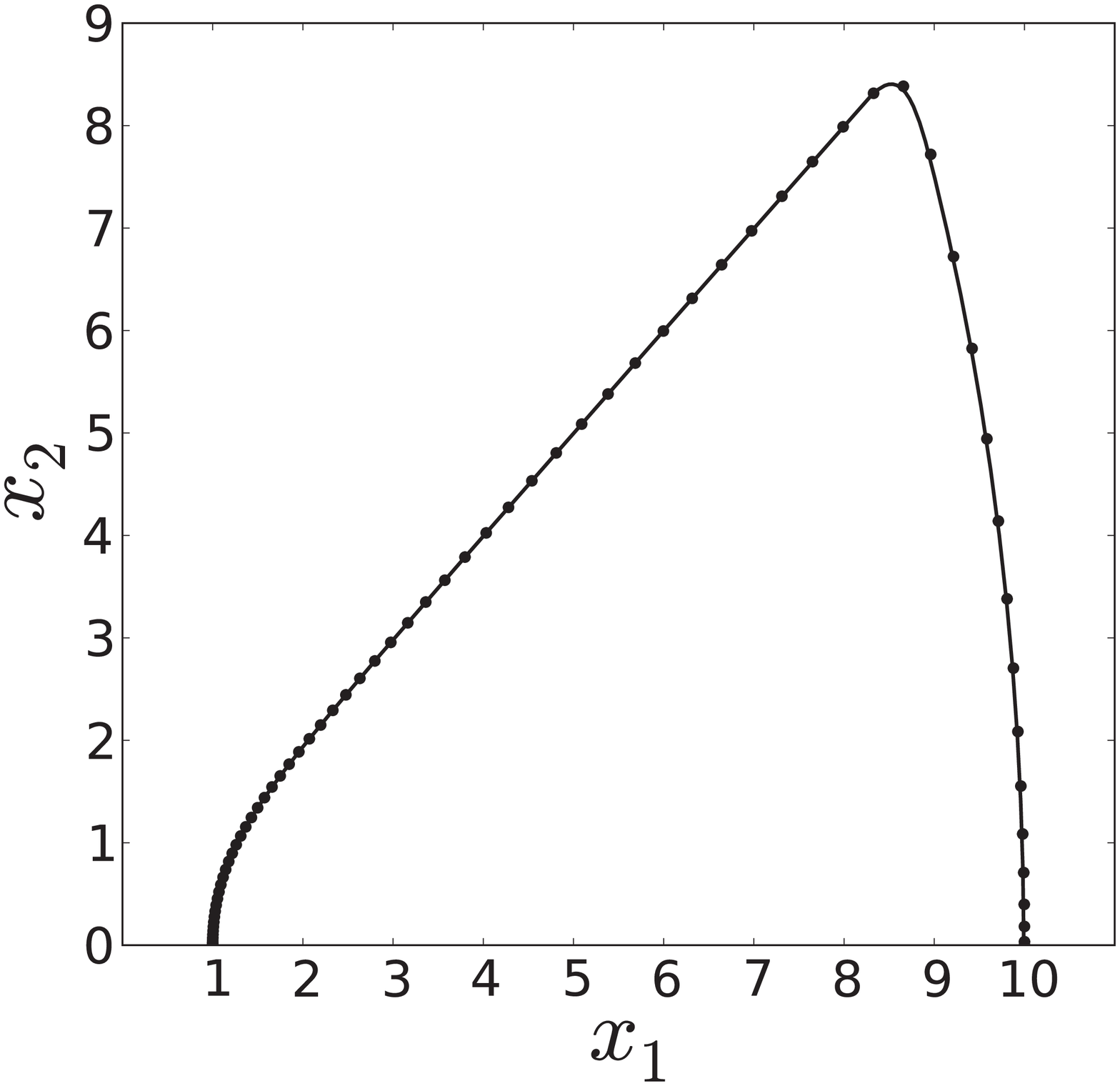}} \\
	        \subfigure[$\ $Realistic control, $M=10$]{
	            \label{fig:ps_sl_con_1}
	            \includegraphics[width=.45\linewidth]{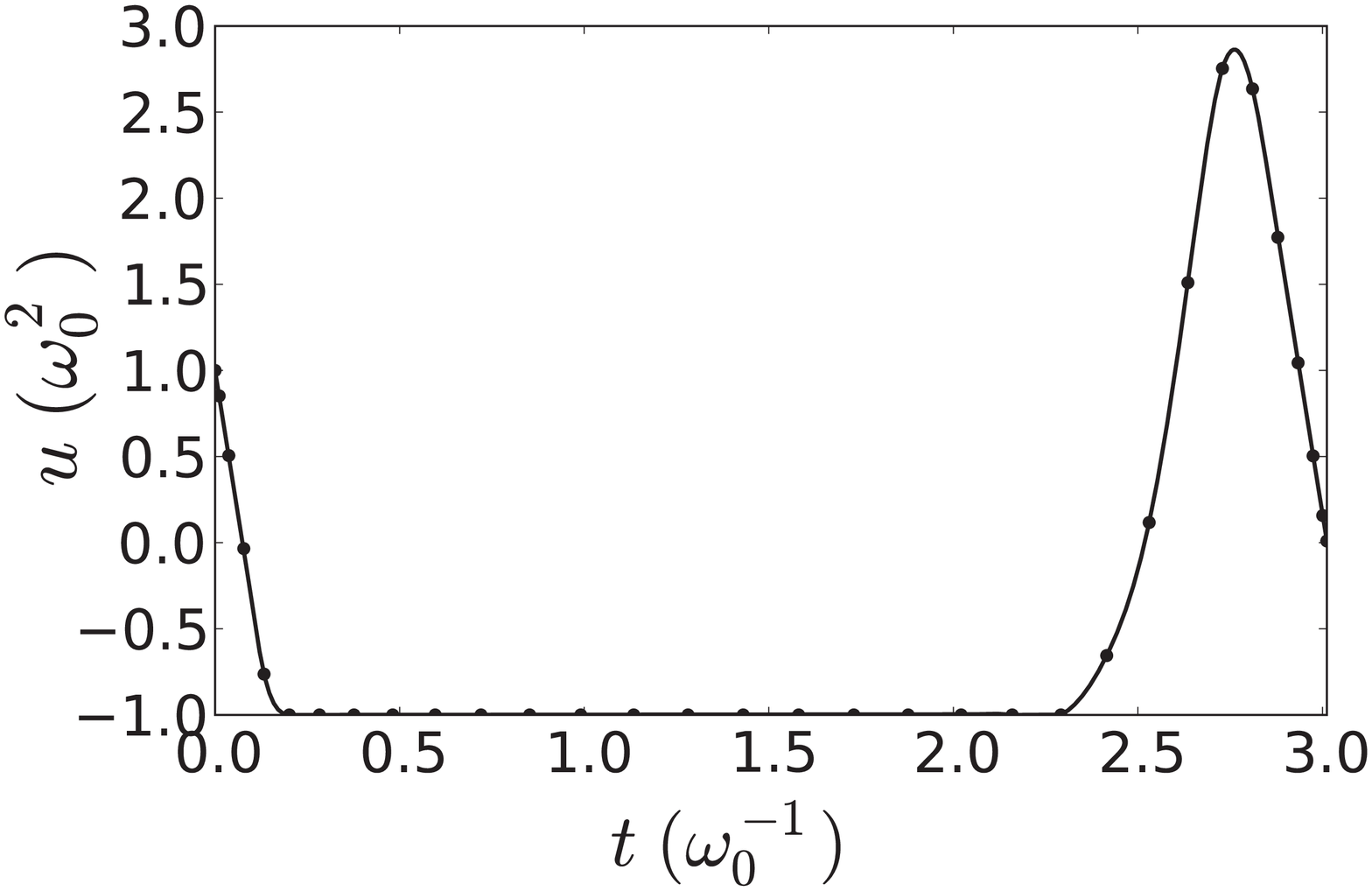}} &
			\subfigure[$\ $Corresponding trajectory]{
	            \label{fig:ps_sl_traj_1}
	            \includegraphics[width=.45\linewidth]{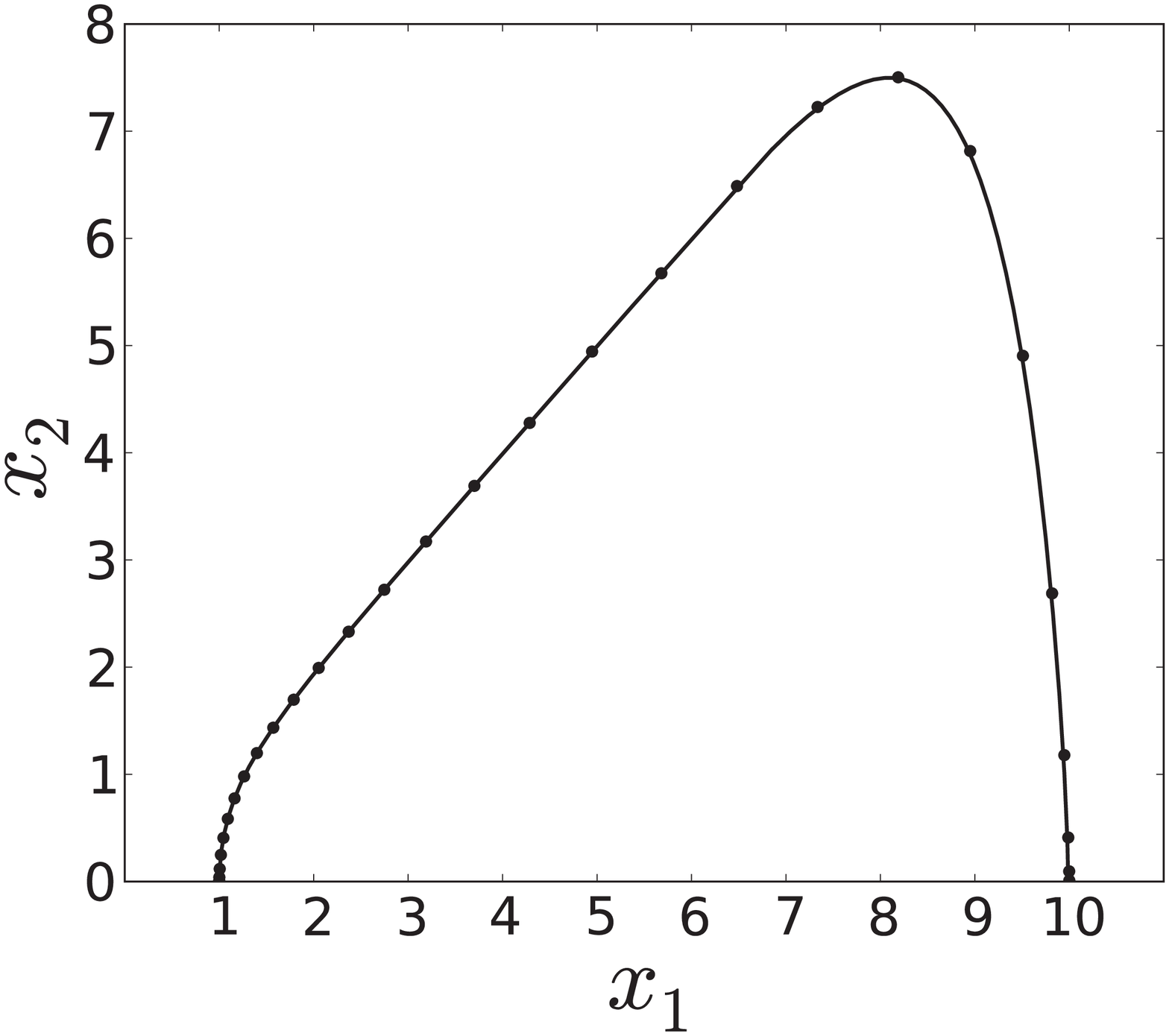}} \\
		\end{tabular}
\caption{Control functions calculated by the pseudospectral method for the same parameters as in Fig. \ref{fig:oneswitching} without ($M=\infty$, panel a) and with ($M=10$, panel c) slope restriction. The latter case requires a larger transfer time, as expected. The corresponding trajectories are also shown (panels b,d).}
 \label{fig:smooth1}
\end{figure}

\begin{figure}[t]
 \centering
		\begin{tabular}{cc}
     	\subfigure[$\ $Optimal control, $M=\infty$]{
	            \label{fig:ps_con_2}
	            \includegraphics[width=.45\linewidth]{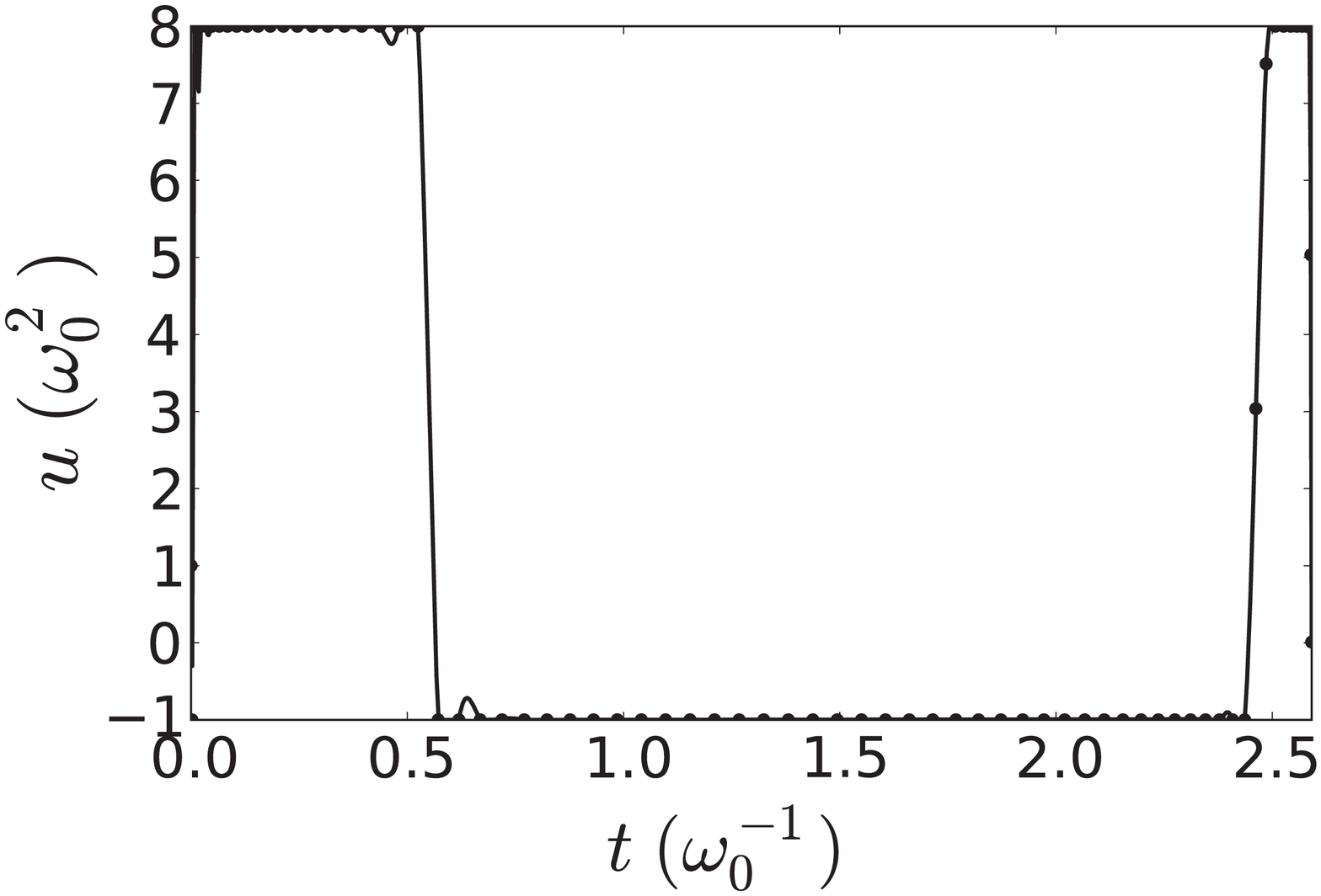}} &
	        \subfigure[$\ $Corresponding trajectory]{
	            \label{fig:ps_traj_2}
	            \includegraphics[width=.45\linewidth]{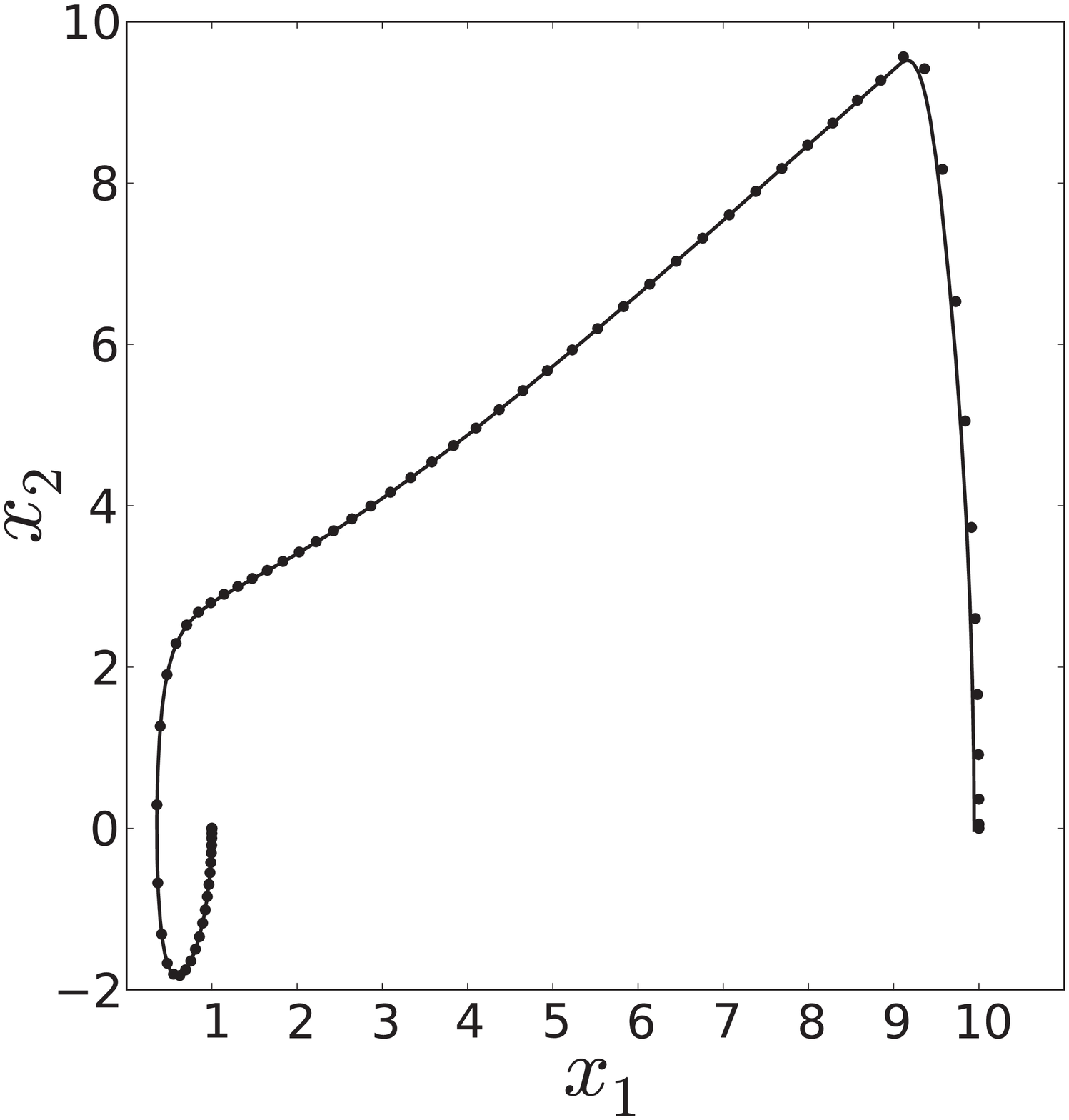}} \\
	        \subfigure[$\ $Realistic control, $M=10$]{
	            \label{fig:ps_sl_con_2}
	            \includegraphics[width=.45\linewidth]{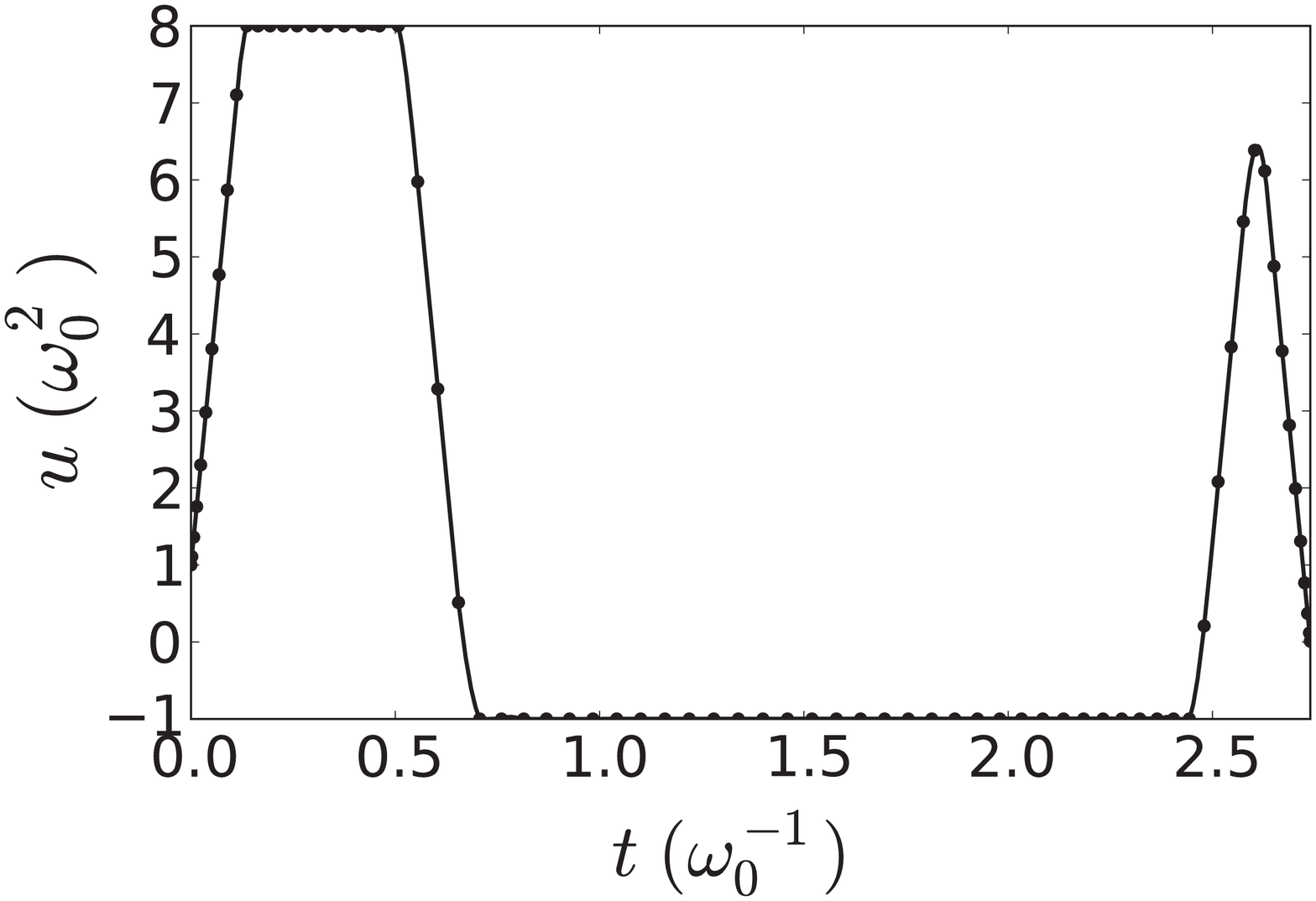}} &
			\subfigure[$\ $Corresponding trajectory]{
	            \label{fig:ps_sl_traj_2}
	            \includegraphics[width=.45\linewidth]{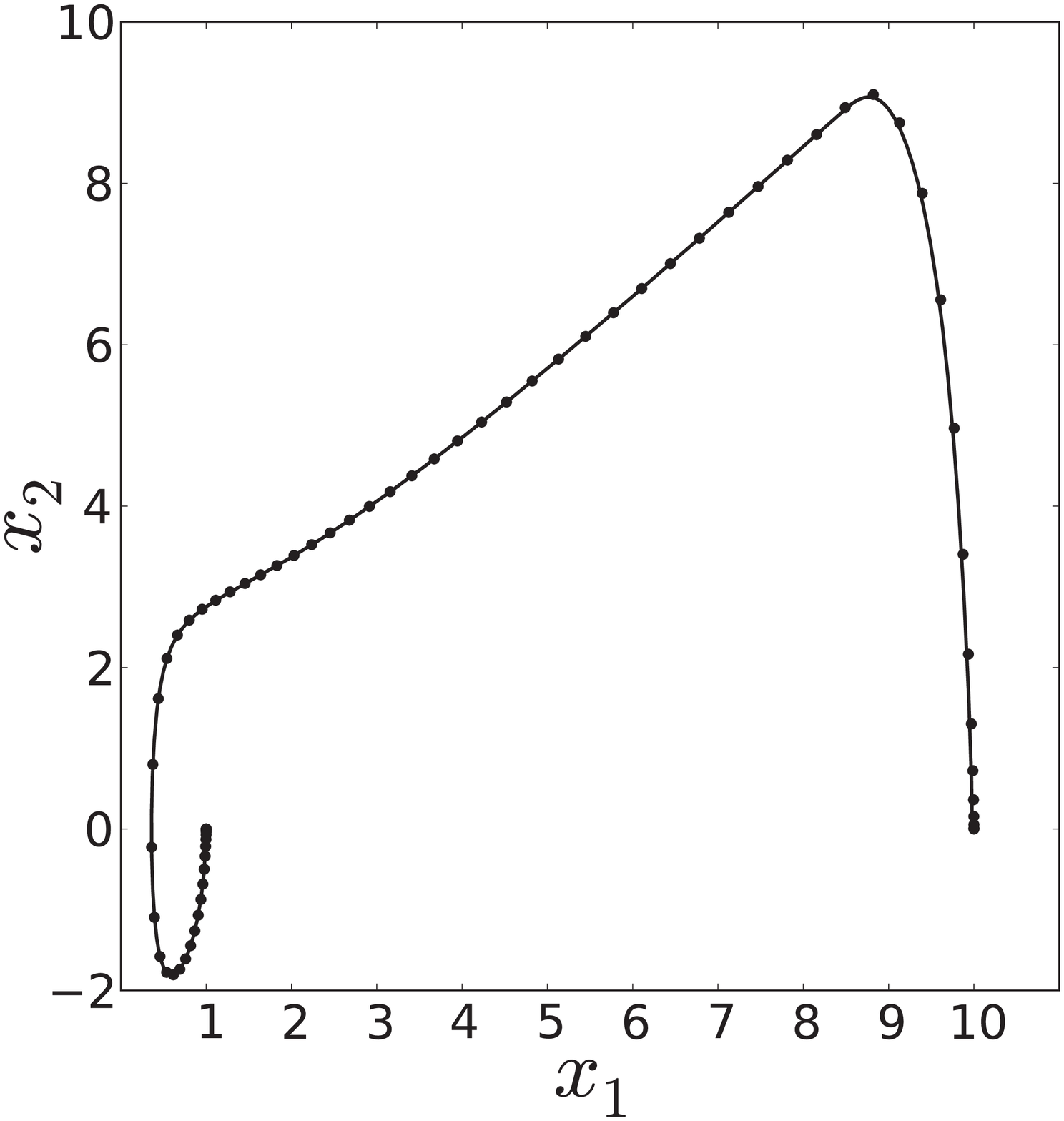}} \\
		\end{tabular}
\caption{Control functions calculated by the pseudospectral method for the same parameters as in Fig. \ref{fig:switchingstwo} without ($M=\infty$, panel a) and with ($M=10$, panel c) slope restriction. The latter case requires a larger transfer time, as expected. The corresponding trajectories are also shown (panels b,d).}
 \label{fig:smooth2}
\end{figure}

For an arbitrary selection of nodes, as the order of approximation $N$ gets large, Runge oscillations near the endpoints of the $[-1,1]$ domain may occur \cite{fornberg_practical_1998}, as shown in Fig. \ref{fig:uniform_lgl}. In order to suppress this phenomenon and increase the accuracy of the approximation, we use the Legendre-Gauss-Lobatto (LGL) nodes, which are the end points $t_0=-1,t_N=1$ and the roots of $\dot{L}_N(t)$, the derivative of the $N^{\rm th}$ order Legendre polynomial \cite{boyd_chebyshev_2000}. The corresponding grid is $\Gamma^{LGL}=\{t_i:t_0=-1,\dot{L}_N(t)|_{t_j}=0, i=1,\ldots N-1,t_N=1\}$. In this case the Lagrange polynomials $\ell_k(t)$ can be expressed as
\begin{equation}\label{eq:lagleg}
	\ell_k(t) = \displaystyle\frac{1}{N(N+1)L_N(t_k)}\frac{(t^2-1) \dot{L}_N(t)}{t-t_k},
\end{equation}
where $\{t_k\}\in\Gamma^{LGL}$, $k=0,1,\ldots,N$.

The $N^{\text{th}}$ order interpolating approximations of the state trajectory and control functions with respect to the same grid are,
\begin{eqnarray} 
	\label{eq:Ix} \mathbf{x}(t) &\approx I_N \mathbf{x}(t) = \sum_{k=0}^N \mathbf{x}_k \ell_k(t), \\
	\label{eq:Iu} u(t) &\approx I_N u(t) = \sum_{k=0}^N u_k \ell_k(t),
\end{eqnarray}
where $\mathbf{x}_k$ and $u_k$ are not only the coefficients of the expansions, but also the function values at the $k^{\text{th}}$ node due to the definition of the Lagrange polynomials \cite{elnagar_pseudospectral_1995}. From the interpolation as in \eqref{eq:Ix}, we have
\begin{equation*}\label{eq:dinterpsc}	
\frac{d}{dt} I_N \mathbf{x}(t) = \sum_{k=0}^N \mathbf{x}_k \dot{\ell}_k(t).
\end{equation*}
Using \eqref{eq:lagleg} and special recursive identities for the derivative of Legendre polynomials \cite{williams_gauss--lobatto_2006}, we have at the LGL nodes $t_i\in\Gamma^{LGL}$, $i=0,1,\ldots,N$,
\begin{equation}\label{eq:dinterpsc_j}
	\frac{d}{dt} I_N \mathbf{x}(t_i) = \sum_{k=0}^N \mathbf{x}_k \dot{\ell}_k(t_i)=\sum_{k=0}^N D_{ik}\mathbf{x}_k,
\end{equation}
where $D_{ik}$ are $ik^{\rm th}$ elements of the constant $(N+1)\times(N+1)$ differentiation matrix $D$ defined by \cite{gottlieb_theory_1984}
\begin{equation}\label{eq:D}
D_{ik} = \left\{
\begin{array}{cl}
    \frac{L_N(t_i)}{L_N(t_k)}\frac{1}{t_i-t_k} & i \neq k \\  & \\
    -\frac{N(N+1)}{4} & i=k=0 \\ & \\
    \frac{N(N+1)}{4} & i=k=N \\ & \\
    0 & \textrm{otherwise}.
\end{array}
\right.
\end{equation}

The pseudospectral method is a collocation method in which the state dynamics is enforced at the LGL nodes. Using \eqref{modsystem1}, \eqref{modsystem2}, \eqref{eq:Ix}, \eqref{eq:Iu} and \eqref{eq:dinterpsc_j}, we obtain the following dynamic constraints
\begin{eqnarray}
\label{constraint1}
\sum_{k=0}^N D_{ik}x_{1k} & = & \frac{t_f}{2}\,x_{2i}\\
\label{constraint2}
\sum_{k=0}^N D_{ik}x_{2k} & = & \frac{t_f}{2}\left(-u_ix_{1i}+\frac{1}{x_{1i}^3}\right)
\end{eqnarray}
for $i=0,1,\ldots,N$, with $\mathbf{x_i}=(x_{1i},x_{2i})^T$. To prevent unrealistic discontinuities in $u(t)$, we impose the following slope restriction
\begin{equation}
\label{slope}
\frac{u_{i+1}-u_i}{t_{i+1}-t_i}\leq M
\end{equation}
for $i=0,1,\ldots,N-1$, where $M$ characterizes the maximum allowed slope of the control function.
The corresponding finite-dimensional constrained minimization problem is to find minimum $t_f$ and $\{u_{i}\}$ with $-v_1\leq u_i\leq v_2$, such that the above algebraic relations and the boundary conditions $(x_{10},x_{20},u_0)=(1,0,1)$, $(x_{1N},x_{2N},u_N)=(\gamma,0,1/\gamma^4)$ are satisfied. Solvers for this type of problems are readily available. In Figs. \ref{fig:smooth1} and \ref{fig:smooth2} we plot the optimal controls and the corresponding trajectories calculated by the pseudospectral method, for the same parameter values as in Figs. \ref{fig:oneswitching} and \ref{fig:switchingstwo}, respectively, with and without slope restriction.

\section{Conclusion}

In this paper we used optimal control theory to show that minimum time frictionless atom cooling in harmonic traps is achieved when the trap frequency changes in a ``bang-bang" manner, even if the trap is allowed to become transiently an expulsive parabolic potential. Using this fact we calculated estimates of minimum cooling times for control strategies with various numbers of frequency jumps. Finally, we employed a pseudospectral optimization method to find realistic solutions without discontinuities, appropriate for experimental implementation. The above results and techniques are not restricted to atom cooling but are applicable to areas as diverse as adiabatic quantum computing \cite{Aharonov07} and finite time thermodynamic processes \cite{Salamon09}.

\section{Acknowledgements}

This work was supported by the NSF under the Career Award \#0747877 and the AFOSR Young Investigator Award \#FA9550-10-1-0146. The authors would like to thank professor Heinz Schaettler for valuable comments.


\begin{thebibliography}{99}


\bibitem{Leanhardt03}
A.E. Leanhardt, T.A. Pasquini, M. Saba, A. Schirotzek, Y. Shin, D. Kielpinski, D.E. Pritchard, and W. Ketterle, Science 301, 1513 (2003).

\bibitem{Bize05}
S. Bize et al., J. Phys. B: At. Mol. Opt. Phys. 38, S449 (2005).

\bibitem{Aharonov07}
D. Aharonov, W. van Dam, J. Kempe, Z. Landau, S. Lloyd, and O. Regev, SIAM J. Comput. 37, 166 (2007).

\bibitem{Salamon09}
P. Salamon, K.H. Hoffmann, Y. Rezek, and R. Kosloff, Phys. Chem. Chem. Phys. 11, 1027 (2009).

\bibitem{Chen10}
X. Chen, A. Ruschhaupt, S. Schmidt, A. del Campo, D. Gu\'{e}ry-Odelin, and J.G. Muga, Phys. Rev. Lett. 104, 063002 (2010).

\bibitem{Lewis69}
H.R. Lewis and W.B. Riesenfeld, J. Math. Phys. 10, 1458 (1969).

\bibitem{Tarn80}
T.-J. Tarn, G. Huang, and J.W. Clark, Mathematical Modelling 1, 109 (1980).

\bibitem{Peirce88}
A. Peirce, M. Dahleh, and H. Rabitz, Phys. Rev. A 37, 4950 (1988).

\bibitem{Khaneja01}
N. Khaneja, R. Brockett, and S.J. Glaser,  Phys. Rev. A 63, 032308 (2001).

\bibitem{D'Alessandro01}
D. D' Alessandro and M. Dahleh, IEEE Trans. Autom. Contr. 46, 866 (2001).

\bibitem{Lloyd01}
S. Lloyd and L. Viola, Phys. Rev. A 65, 010101(R) (2001).

\bibitem{Sklarz02}
S.E. Sklarz and D.J Tannor, Phys. Rev A 66, 053619 (2002).

\bibitem{Boscain02}
U. Boscain, G. Charlot, J.P. Gauthier, S. Guerin, and H.R. Jauslin, J. Math. Phys. 43, 2107 (2002).

\bibitem{Skinner03}
T.E. Skinner, T. Reiss, B. Luy, N. Khaneja, and S. J. Glaser, J. Magn. Reson. 163, 8 (2003).

\bibitem{Stefanatos04}
D. Stefanatos, N. Khaneja, and S. J. Glaser, Phys. Rev. A 69, 022319 (2004).

\bibitem{Stefanatos05}
D. Stefanatos, S. J. Glaser, and N. Khaneja, Phys. Rev. A 72, 062320 (2005).

\bibitem{Li06}
J.-S. Li and N. Khaneja, Phys. Rev. A 73, 030302(R) (2006).

\bibitem{Gorshkov08}
A.V. Gorshkov, T. Calarco, M.D. Lukin, and A.S. S{\o}rensen, Phys. Rev. A 77, 043806 (2008).

\bibitem{Maximov08}
I.I. Maximov, Z. Tošner, and N.C. Nielsen, J. Chem. Phys. 128, 184505 (2008).

\bibitem{Li09}
J.-S. Li, J. Ruths, and D. Stefanatos, J. Chem. Phys. 131, 164110 (2009).

\bibitem{Liieee09}
J.-S. Li, N. Khaneja, Ensemble Control of Bloch Equations, IEEE Trans. Autom. Control 54, 528 (2009).

\bibitem{Wu09}
R. Wu, J. Dominy, T.-S Ho, and H. Rabitz (2009), arXiv:0907.2354 [quant-ph].

\bibitem{Lapert10}
M. Lapert, Y. Zhang, M. Braun, S. J. Glaser, and D. Sugny, Phys. Rev. Lett. 104, 083001 (2010).

\bibitem{Schulte10}
T. Schulte-Herbr\"{u}ggen, S.J. Glaser, G. Dirr, and U. Helmke, Rev. Math. Phys. 22, 597 (2010).

\bibitem{Stefanatos10}
D. Stefanatos and J.-S. Li, Syst. Contr. Lett. (to be published).

\bibitem{Pontryagin}
L.S. Pontryagin, V.G. Boltyanskii, R.V. Gamkrelidze, and E.F.
Mishchenko, {\it The Mathematical Theory of Optimal Processes} (Interscience Publishers, New York, 1962).

\bibitem{elnagar_pseudospectral_1995}
G. Elnagar, M.A. Kazemi, and M. Razzaghi, IEEE Trans. Autom. Contr. 40, 1793 (1995).

\bibitem{ross_legendre_2004}
I. Ross and F. Fahroo, in {\it New Trends in Nonlinear Dynamics and Control}, edited by W. Kang et al. (Springer, Berlin, 2003), p. 327.

\bibitem{wei_kang_convergence_2006}
W. Kang, Q. Gong, and I. Ross, IEEE Trans. Autom. Contr 51, 1115 (2006).

\bibitem{fahroo_costate_2001}
F. Fahroo and I. Ross, J. Guid. Contr. Dynam. 24, 270 (2001).

\bibitem{williams_gauss--lobatto_2006}
P. Williams, ANZIAM 47, C101 (2006).

\bibitem{canuto_spectral_2006}
C. Canuto, M.Y. Hussaini, A. Quarteroni, and T.A. Zang, {\it Spectral Methods} (Springer, Berlin, 2006).

\bibitem{davis_interpolation_1963}
P.J. Davis, {\it Interpolation and Approximation} (Blaisdell, New York, 1963).

\bibitem{szego_orthogonal_1959}
G. Szego, {\it Orthogonal Polynomials} (American Mathematical Society, New York, 1959).

\bibitem{fornberg_practical_1998}
B. Fornberg, {\it A Practical Guide to Pseudospectral Methods} (Cambridge University Press, New York, 1998).

\bibitem{boyd_chebyshev_2000}
J. Boyd, {\it Chebyshev and Fourier Spectral Methods} (Dover Ed. 2, New York, 2000).


\bibitem{gottlieb_theory_1984}
D. Gottlieb, Y. Hussaini, and S. Orszag, in {\it Spectral Methods for Partial Differential Equations}, edited by R. Voigt et al. (SIAM, Philadelphia, 1984), p. 1.

\end{thebibliography}

\end{document}